\documentclass[11pt,a4paper]{article}
\usepackage{jheppub}
\usepackage{amsmath}
\usepackage{multicol,bbm}
\usepackage{enumerate}
\usepackage{bibentry}

\graphicspath{{./Figs/}}

\def\be{\begin{equation}}
\def\ee{\end{equation}}
\def\ba{\begin{eqnarray}}
\def\ea{\end{eqnarray}}
\def\bea{\begin{eqnarray}}
\def\eea{\end{eqnarray}}
\def\eq{\begin{equation}}
\def\eqe{\end{equation}}
\def\eqa{\begin{eqnarray}}
\def\eqae{\end{eqnarray}}

\def\p{\partial}

\title{On the exactness of soft theorems}
\author[a,b, c]{Andrea L. Guerrieri,}
\author[d,e]{ Yu-tin Huang,}
\author[d]{Zhi-Zhong Li,}
\author[f, g]{Congkao Wen}
\affiliation[a]{ICTP South American Institute for Fundamental Research, IFT-UNESP, S\~ao Paulo, SP Brazil  01440-070} 
\affiliation[b]{
Department of Physics, Faculty of Science, Chulalongkorn University, Thanon \\Phayathai,
Pathumwan, Bangkok 10330, Thailand
}
\affiliation[c]{
I.N.F.N. Sezione di Roma Tor Vergata, Via della Ricerca Scientifica 00133 Roma, Italy
}
\affiliation[d]{Department of Physics and Astronomy, National Taiwan University, Taipei 10617, Taiwan}
\affiliation[e]{Physics Division, National Center for Theoretical Sciences, National Tsing-Hua University,
No.101, Section 2, Kuang-Fu Road, Hsinchu, Taiwan} 
\affiliation[f]{Walter Burke Institute for Theoretical Physics, California Institute of Technology, Pasadena, CA, 91125, USA}
\affiliation[g]{Mani L. Bhaumik Institute for Theoretical Physics, Department of Physics and Astronomy,
UCLA, Los Angeles, CA 90095, USA}
\emailAdd{aguerrieri@ictp-saifr.org, yutinyt@gmail.com, b02202003@ntu.edu.tw, cwen@caltech.edu} 

\abstract{Soft behaviours of S-matrix for massless theories reflect the underlying symmetry principle that enforces its masslessness. As an expansion in soft momenta, sub-leading soft theorems can arise either due to (I) unique structure of the fundamental vertex or (II) presence of enhanced broken-symmetries. While the former is expected to be modified by infrared or ultraviolet divergences, the latter should remain exact to all orders in perturbation theory. Using current algebra, we clarify such distinction for spontaneously broken (super) Poincar\'e and (super) conformal symmetry.  We compute the UV divergences of DBI, conformal DBI, and A-V theory to verify the exactness of type (II) soft theorems, while type (I) are shown to be broken and the soft-modifying higher-dimensional operators are identified. As further evidence for the exactness of type (II) soft theorems, we consider the $\alpha'$ expansion of both super and bosonic open strings amplitudes, and verify the validity of the translation symmetry breaking soft-theorems up to $\mathcal{O}(\alpha'^{6})$. Thus the massless S-matrix of string theory ``knows" about the presence of D-branes. }
\preprint{CALT-TH-2017-26, NCTS-TH/1709}

\begin{document}
\maketitle 
\section{Introduction and motivations}
Soft behaviour of the S-matrix for massless theories, where one considers an expansion in small momenta for one or more external legs, exhibits universal behaviour that reflects the underlying symmetry principle. Indeed the Ward identity from gauge invariance directly leads to Weinberg's soft theorem~\cite{Weinberg:1965nx}, while spontaneous broken symmetry (SBS) is reflected in Adler's zero~\cite{Adler:1964um}. In other words, soft theorems are the on-shell avatar of the symmetry constraints that protect the light degrees of freedom from quantum corrections. 

Universality of the sub-leading expansion can arise from two origins. (I) First, the symmetry principle that leads to the leading soft theorem will have sub-leading extensions when combined with the specific structure of the lowest multiplicity vertex. More precisely, focusing on the factorization diagrams with one fundamental vertex on one-side will give the leading soft-limit behaviour. For sub-leading, although other diagrams also contributes, the underlying symmetry principle can relate them to that of the factorization diagram. Indeed this precisely was how the sub-leading photon and graviton soft theorems were derived initially~\cite{Low:1958sn, Burnett:1967km, GellMann:1954kc,Gross:1968in, Bern:2014vva}, and recently extended to SBS ~\cite{Low:2015ogb}. For these theories, one can show that soft theorems are sufficient to determine the full tree-level S-matrix via recursion relations~\cite{Cheung:2015ota, Luo:2015tat}.\footnote{In fact in some cases, the S-matrix is fixed simply by simultaneously impose soft theorems and locality constraints~\cite{Arkani-Hamed:2016rak, Rodina:2016mbk, Rodina:2016jyz}.}  More generally, soft theorems play important roles in constraining the low-energy effective actions~\cite{Bianchi:2016viy} (also see~\cite{Cheung:2016drk}), especially when combine with constraints from supersymmetry~\cite{Chen:2015hpa}. The fact that soft-contributions factorizes is also crucial in the exponentiation of soft emissions \cite{Yennie:1961ad} in gauge theories, and has also been extended to subleading corrections (see \cite{Laenen:2010uz} for a comprehensive review).

(II) Second, if the Goldstone mode is associated with more than one broken generators (``enhanced" broken symmetry), which occurs when the currents are derivatively related, then the linear relation amongst the currents implies universal sub-leading soft behaviours.  More precisely, consider the case where there is a set of broken generators $\{G_i\}$, where the algebra admits the following commutation relation:
\eq
[P, G_1]\sim G_2\,.
\eqe
Here $P$ is the translation generator. The Goldstone mode associated with broken generator $G_2$ is then derivatively related to that of $G_1$, and hence they should be identified. Since the translation generator $P$ is involved in the algebra, such scenarios occur for spontaneously broken space-time symmetries. In the language of currents, one would schematically have $J_1 \sim x J_2$, which in momentum space becomes 
\eq
\tilde{J}_1 \sim \frac{\partial}{\partial p} \tilde{J}_2\,,
\eqe
where $\tilde{J}$'s are the Fourier transformed currents. Thus as one applies Ward identities to derive soft theorems, using $\tilde{J}_1$ instead of $\tilde{J}_2$ to excite the soft Goldstone mode would lead to soft theorems that are sub-leading in the soft-momenta expansion. In other words, the presence of sub-leading soft theorems reflect the enhanced broken symmetry associated with the single Goldstone boson. Examples of such symmetry breaking pattern include conformal symmetry and translation symmetry, where one identifies $\{G_1,G_2\}$ as $\{ K,D\}$ and $\{\hat{L},\hat{P}\}$ respectively~\cite{Low:2001bw},\footnote{Here the hat indicates these are translation and rotation symmetries involving extra-dimensional directions that are spontaneously broken.} with $\hat{L}$ the angular momentum and $\{K,D\}$ being the special conformal transformation and dilation generators. Indeed sub-leading single-soft theorems for the dilaton were derived using the above approach in~\cite{DiVecchia:2015jaq}. In short, type (I) soft theorems depends on both the symmetry and detailed structure of the interaction vertex, whether it is tree-level or loop-level generated counter terms. Type (II) is determined by symmetry alone.

Type (I) soft theorems are not expected to survive quantum corrections simply due to the fact that IR or UV divergences can modify the structure of the fundamental vertex. Indeed sub-leading graviton and photon soft theorems are modified at loop-level due to IR divergences~\cite{Bern:2014vva, He:2014bga}, or UV divergences that introduce higher dimensional three-point operators~\cite{Bianchi:2014gla}. The modifications due to higher-dimensional operators were systematically analysed in~\cite{Elvang:2016qvq}, where the relevant operators as well as the resulting modified soft theorems were identified.

Since type (II) sub-leading soft theorems can be derived using symmetry principle based current algebra, if the symmetry is not anomalous, it should hold to all orders and irrespective of the details of the UV completion. The aim of this paper is two folds:
\begin{itemize}
  \item Deriving sub-leading soft theorems from current algebra for spacetime symmetry breaking. More precisely we derive the single- and double-soft theorems associated with spontaneous conformal as well as Poincar\'e symmetry breaking. As one of the consequences, we clarify which of sub-leading soft-theorems that were derived in~\cite{Cachazo:2015ksa}, for Dirac-Born-Infeld (DBI) are of type II. 
  \item Via explicit computation of UV divergences for various effective field theories, we verify that type II soft theorems  survive quantum corrections. Note that this closes a tiny loop-hole in the discussion of counter terms in supergravity, where their compatibility with duality symmetries are analysed via soft theorems~\cite{ArkaniHamed:2008gz, Elvang:2010kc, Beisert:2010jx}. In principle one has to show that the regulated theory leaves the duality intact, which is not trivial since the duality involves the electric magnetic duality of the photon fields, which is defined strictly in four-dimensions.  Through maximal susy, the duality symmetry is directly related to the scalar sector where the duality implies explicit double-soft theorems. 
   \end{itemize}

We begin by re-deriving the double soft theorems associated with translational symmetry up to order $\tau$, where $\tau$ parameterises the soft momenta $p_i$ by $p_i \rightarrow \tau p_i$,  and conjecture that the $\mathcal{O}(\tau^2)$ also has symmetry based origin. Similarly,  double soft theorems for spontaneously broken conformal symmetry are derived up to the same order. This result indicates that $\mathcal{O}(\tau^3)$ soft theorem found in~\cite{Cachazo:2015ksa} for the DBI action cannot be respected by its UV divergences. As a check, we compute the one-loop UV divergences up to six points for DBI, conformal DBI\footnote{Here by conformal DBI we mean the DBI action in a AdS background.} as well as that for the Akulov-Volkov (A-V) theory~\cite{Volkov:1972jx, Volkov:1973ix}, namely the effective action for spontaneously supersymmetry breaking. For the former, we have verified that indeed the $\mathcal{O}(\tau^1)$ and $\mathcal{O}(\tau^2)$ soft theorems are respected, while the $\mathcal{O}(\tau^3)$ soft-theorem is indeed broken. We have further identified the culprit of this violation to be the presence of a new eight-derivative counter term in four-dimensions. In six-dimensions such counter term does not arise, and the $\mathcal{O}(\tau^3)$ soft-theorem is restored. This confirms the claim that $\mathcal{O}(\tau^3)$ soft-theorems are due to the structure of the tree-level four-point vertex. Similarly we verified that the leading single- and double-soft theorems are also respected by the UV divergences of A-V theory. We note that this is the first time the double-soft theorems associated with non-linear symmetries are tested against non-trivial UV divergences.

The fact that type (II) soft-theorems survive quantum corrections imply that it should apply to any UV completion. As a test, we consider the super and bosonic open string theories which are the UV completions of the effective field theory of D-branes in flat space. We consider the scalar modes that come from dimensionally reduced amplitudes. Since this system corresponds to the Goldstone modes of D-branes, we should expect that the single and double-soft theorems due to the (broken) translation symmetry are respected. We verify this against the four and six-point abelianized open string amplitudes\footnote{The lowest order in $\alpha'$-expansion of abelianized open string amplitudes corresponds to the amplitudes in DBI.}, and show that once again the double-soft theorems are respected up to $\mathcal{O}(\tau^2)$, while the $\mathcal{O}(\tau^3)$ is broken. Thus the S-matrix of string theory knows about the presence of D-branes. Note that since the previous analysis shows that eight-derivative four-point operators modify the $\mathcal{O}(\tau^3)$ soft theorem, by power counting one would argue that the presence of six-derivative counter terms would modify $\mathcal{O}(\tau^2)$ soft theorems. Indeed such operator does appear in the bosonic string effective action. The fact that  $\mathcal{O}(\tau^2)$ soft theorem is respected substantiate our claim that it is symmetry protected. Finally, soft-theorems are related to the degeneracy of the vacuum manifold, it would hold even if the vacuum is unstable, as is the case for the bosonic string.

This paper is organised as follows: In the next section~\ref{sec:derivation}, we review the derivation of soft theorems from Ward identities of currents. We will argue that for degenerate currents, in that one is related to the derivative of the other, soft theorems for the Goldstone bosons can be derived up to order $\mathcal{O}(\tau)$ for the single soft limit, and $\mathcal{O}(\tau^2)$ for the double soft limit. In section~\ref{sec:UV-divergence}, we compute the one-loop UV divergences up to six points in the theories of DBI, conformal DBI as well as A-V theory, where we demonstrate that the symmetry derived soft theorems are respected, while those derived in~\cite{Cachazo:2015ksa} that are not, are broken by loop-level corrections. In section~\ref{sec:strings}, we study massless amplitudes in open string theory, and consider the dimensionally reduced amplitude where the modes are Goldstone bosons. We show that the previously derived soft theorems are again satisfied. Finally, in section~\ref{sec:conclusion} we present our conclusions and outlook. Some technical issues are discussed in the appendices. 

\vspace*{5mm}

{\bf Note Added:} \textit{In the final stages of this draft, the work of Paolo Di Vecchia, Raffaele Marotta, and Matin Mojaza appeared on arXiv~\cite{DiVecchia:2017uqn}, where the conformal double-soft theorems were also derived with similar methods. }


\section{Derivation of soft theorems}
\label{sec:derivation}
Here we review the derivation of soft theorems from current algebra, where we use the currents associated with the broken symmetries to excite the Goldstone boson. Let us begin with the Ward Identity:\footnote{We consider the Ward identity of $j_D$ first because it makes our calculation simpler.}
\eqa\label{eq1}
&&\int d^Dx\, e^{iq\cdot x}\prod_i\left[ \int d^Dx_i \, e^{ik_i\cdot x_i}\partial^2_i\right] \frac{\partial}{\partial x^\mu}  \langle 0 | j_{A}^\mu (x) j_{A}^{\mu_1} (x_1) \cdots j_{A}^{\mu_n} (x_n) | 0 \rangle
\\
&=& \sum_i \int d^Dx_i\, e^{i(k_i+q)\cdot x_i}\partial^2_i\,\prod_{m\neq i}\left[ \int d^Dx_m e^{ik_m\cdot x_m}\partial^2_m\right]\displaystyle\sum_i \langle 0 | j_{A}^{\mu_1} (x_1) \cdots \delta_A j_{A}^{\mu_i} (x_i) \cdots j_{A}^{\mu_n} (x_n) | 0 \rangle \nonumber
\eqae
where we Fourier transform one of the legs with momentum $q$, while apply LSZ reduction on the remaining ones. We will use the subscript $A$ in $j_{A}^{\mu_i}$ to indicate it is the current of a broken generator $j_{A}^{\mu_i}|0\rangle\neq0$. The broken current excites a Goldstone boson from the vacuum, $\langle \pi(p)|j_A^{\mu}(x)|0\rangle\sim F p^\mu e^{ip\cdot x}$ and thus in a correlator one will find
\eq\label{ward}
\langle 0 | j_{A}^\mu \cdots | 0 \rangle= \int_p \langle 0 | j_{A}^\mu|\pi\rangle\frac{1}{p^2}\langle\pi| \cdots | 0 \rangle=\int_p Fe^{ip\cdot x}\frac{ p^\mu }{p^2}\langle\pi| \cdots | 0 \rangle
\eqe
where $\langle \pi|$ is the Goldstone boson interpolating field whose propagator is $\frac{1}{p^2}$. The first line of eq.(\ref{eq1}) in the limit $q\rightarrow 0$ becomes
\eq
M(\pi_q\pi_1\cdots \pi_n)\prod_{i=1}^np^{\mu_i}_i+\mathcal{O}(q^1)\,.
\eqe
The second line now depends on whether $\delta_A j_{A}^{\sigma_i}$ produces a state in the physical spectrum. For theories that do not produce such states,  the result is $0$ due to the projection of LSZ reduction. This is the famous Adler's zero. This is indeed the case for Non-Linear Sigma Models, since 
\eq
\delta_{A^a_1} j_{A^{a2}}=f^{a_1a_2b_1} j_{V^{b1}}
\eqe 
where $a_i$ label the distinct generators and $j_{V^{b1}}$ is a current of the unbroken invariant subgroup and hence does not produce physical states in the spectrum, i.e. $j_{V}^{\mu_i}|0\rangle=0$.

For cases that it does produce physical states, then the single-soft limit no-longer is zero.  A prime example of the latter case is spontaneous broken conformal symmetry. First of all, although both dilatation and conformal boost symmetries are broken, there is only one Goldstone boson. This is because 
\eq
[P,K]\sim D
\eqe
and thus the Goldstone mode from $K$ is derivatively related to that of $D$~\cite{Low:2001bw}. Indeed this relation can be realised on the explicit form of the corresponding currents:
\eq
j^\mu_{D}=T^{\mu\nu}x_\nu, \quad j^\mu_{K^\nu}=T^{\mu\lambda}(2x_\nu x_\lambda-\delta_{\nu\lambda}x^2)
\eqe
where $T^{\mu\nu}$ is the stress tensor. Now since the current itself also transforms linearly under both $K$, and $D$, the RHS of eq.~(\ref{eq1}) no longer vanishes, i.e. one has non-zero soft limits.

Furthermore, the fact that the two currents associated with the same Goldstone boson are derivatively related implies that soft theorems can be extended to the sub-leading level. To see this, first note that:
\eqa
\p_\mu \langle j_D^\mu \cdots \rangle&=&\langle T_\mu^\mu \cdots \rangle + x_\nu \p_\mu \langle T^{\mu \nu} \cdots \rangle \, , \nonumber\\
\p_\mu \langle j_{K^\lambda}^\mu \cdots \rangle&=&2x_\lambda \langle T_\mu^\mu \cdots \rangle + (2x_\nu x_\lambda-\eta_{\rho\nu}x^2) \p_\mu \langle T^{\mu \nu} \cdots \rangle \, .
\eqae
Now, since upon Fourier transform in momentum space $\p_{\mu}\langle T^{\mu\nu} \cdots \rangle$ generates a sum over the momenta of the remaining fields, this term will not contribute. Thus we can effectively equate:
\begin{equation}\label{KDRel}
\p_\mu \langle j_{K^\lambda}^\mu \cdots \rangle \sim 2x_\lambda \p_\mu \langle j_D^\mu \cdots \rangle \, .
\end{equation}
Now consider the case where one uses $j^\mu_{K^\nu}$ in eq.~(\ref{eq1}) instead of $j^\mu_{D}$, then due to the extra factor of $x$ in the RHS of eq.~(\ref{KDRel}) the LHS of eq.~(\ref{eq1}), again in the limit $q\rightarrow0$ yields:
\eq
\frac{\partial}{\partial q}M(\pi_q\pi_1\cdots \pi_n)\prod_{i=1}^np^{\mu_i}_i+\mathcal{O}(q^1)\,.
\eqe
This would then lead to soft theorems sub-leading in the soft momenta expansion. 

For the double soft limits, one instead begins with    
\eqa\label{eq2}
&&\prod_i\left[ \int d^Dx_i e^{ik_i\cdot x_i}\partial^2_i\right]\int d^Dy\, e^{ip\cdot y}\frac{\partial}{\partial y^\nu}\int d^Dx\, e^{iq\cdot x} \frac{\partial}{\partial x^\mu} \langle 0 | j_{A}^\mu (x)j_{A'}^\nu (y) \cdots | 0 \rangle\,. 
\eqae
The above equation can be evaluated in two ways. First, integrating by parts both the derivatives in $x$ and $y$ and taking the momenta $p, q$ to be soft, one obtains the double soft limit of $(n{+}2)$-point amplitude. On the other hand, one can also employ the Ward identity, which generates transformations on the other fields in the correlation function. Similar to the single-soft discussion, the resulting double-soft limit depends on the nature of the broken symmetry. There are two sources for non-vanishing results. The first is similar to the single-soft limit, where one considers the variation of remaining fields under the broken symmetries. The second is the variation of the current itself under the broken symmetry, which will be proportional to the current of either an unbroken or broken symmetry. Non-linear sigma models as well as broken translational and supersymmetry are of the former. Broken (super) conformal will receive contributions for both cases. Finally as with the single soft discussion, employing $j^\mu_{K^\nu}$ instead of $j^\mu_{D}$ will lead to sub-leading soft theorems. 

In the following, we will perform a detailed analysis for two types of space-time symmetry breaking, conformal as well as translation. For completeness, we will also derive the double soft theorems for susy and conformal susy breaking in appendix~\ref{Qbreak} and appendix~\ref{Sbreak}.

\subsection{Broken conformal symmetry}
We begin by reviewing the results of the single soft limits of broken conformal symmetry, which is given by, 
\eqa \label{eq:softlimitCDBI}
M_{n+1}{\big |}_{p_{n+1} \rightarrow 0} =  
\left(  \mathcal{S}^{(0)} + \mathcal{S}^{(1)} \right) M_{n}(p_1 , {\cdots}, p_{n}) + \mathcal{O}(p_{n+1}^2) \, .
\eqae
The leading~\cite{Boels:2015pta,Huang:2015sla} and sub-leading~\cite{DiVecchia:2015jaq} soft operators $\mathcal{S}^{(0)}, \mathcal{S}^{(1)}$ are given as
\eqa\label{eq:softnomassS1}
\mathcal{S}^{(0)} &=&  -\sum^{n}_{i=1}
 \left( p_i \cdot {\p \over \p p_i } + {D- 2 \over 2} \right) + D \, ,\cr
\mathcal{S}^{(1)}  &=&  - p^{\mu}_n \sum^{n}_{i=1} 
\left[   p^{\nu}_i { \p^2 \over \p {p_i^{\nu}} \p {p_i^{\mu}} }
-
{\frac{p_{i\mu}}{2}}{ \p^2 \over \p {{p_i}_{\nu}} \p {p_i^{\nu}} } 
+ {D-2 \over 2} {\p \over \p {p^{\mu}_i} } \right] \, ,
\eqae
where $D$ is the space-time dimension, and here we only consider the form of the soft theorem on massless degrees of freedom. We will not present the derivation of this single-soft theorem here which was done in~\cite{DiVecchia:2015jaq}, but just to remark that as discussed in the above review, the presence of sub-leading single-soft theorem is related to the fact that the generators of the broken symmetries being derivatively related. 
\subsubsection{Leading double-soft theorems from $(j_{D}, j_{D}) $}
We now directly move to the double soft limit. For the leading order, we consider the double Ward identity for two dilatation currents:
\eqa\label{b}
&&{\rm LSZ}\int d^Dy\, e^{ip\cdot y}\frac{\p}{\p y^\nu}\int d^Dx\, e^{iq\cdot x} \frac{\p}{\p x^\mu} \langle 0 | j_{D}^\mu (x)j_{D}^\nu (y) \phi (x_1) \cdots \phi (x_n) | 0 \rangle\nonumber\\
&=&{\rm LSZ}\int d^Dy\, (-ip_\nu)\bigg[ e^{i(q+p)\cdot y} \langle 0 | \delta_D j_{D}^\nu (y) \phi (x_1) \cdots \phi (x_n) | 0 \rangle \nonumber\\
&&+\sum_{i=1}^ne^{i(q\cdot x_i+p\cdot y)}\langle 0 |  j_{D}^\nu (y) \phi (x_1)\cdots \delta_D\phi(x_i) \cdots \phi (x_n) | 0 \rangle\bigg]\,
\eqae
where for abbreviation ${\rm LSZ} = \prod_i\left[ \int d^Dx_i e^{ik_i\cdot x_i}-\p^2_i\right]$, as the LSZ reduction. In the above we first apply the Ward identity associated with $\partial_x$, and partial integrate $\partial_y$. For the second term in eq.~(\ref{b}) one can directly apply a second Ward identity in momentum space:
\eqa
&&{\rm LSZ}\int d^Dy\, (-ip^\nu)\left[\sum_i e^{i(q\cdot x_i+p\cdot y)}\langle 0 |  j_{D}^\nu (y) \phi (x_1)\cdots \delta_D\phi(x_i) \cdots \phi (x_n) | 0 \rangle\right]\nonumber\\
&=&{\rm LSZ} \, \sum_{i, j}e^{i(q\cdot x_i+p\cdot x_j)}\langle 0 | \phi (x_1)\cdots\delta_D\phi (x_j)\cdots \delta_D\phi(x_i) \cdots \phi (x_n) | 0 \rangle\,
\eqae
and similarly for the first term once one symmetrises $(p\leftrightarrow q)$.  Now use that 
\eq \label{d}
\delta_D\mathcal{O}(x)=(d+x\cdot \p_x)\mathcal{O}(x)\,,
\eqe
where $d$ is the scaling dimension of the operator $\mathcal{O}$, we get 
\eqa
\text{eq.}~(\ref{b})&=&{\rm LSZ} \, \bigg[{1 \over 2}\left(d-D+1-(p+q)\cdot \p_p\right)\sum_j e^{i(p+q)\cdot x_j} \langle 0 | \phi (x_1) \cdots \delta_D\phi(x_j)\cdots\phi (x_n) | 0 \rangle\nonumber\\
&&+\sum_{i, j} e^{i(q\cdot x_i+p\cdot x_j)}\langle 0 | \phi (x_1)\cdots(d_j+x_j\cdot \p_j)\phi (x_j)\cdots (d_i+x_i\cdot \p_i)\phi(x_i) \cdots \phi (x_n) | 0 \rangle\bigg]\nonumber\\
&=&\left(\prod_i k_i^2\right)\bigg[{1 \over 2}(d-D+1)+\sum_i\mathfrak{D}_i\bigg]\sum_j \mathfrak{D}_j \langle 0 | \tilde{\phi}_1\cdots \tilde\phi_n | 0 \rangle+\mathcal{O}(p,q),
\eqae
with following definitions,
\eq
\tilde\phi_i=\tilde\phi(k_i),\quad \mathfrak{D}_i=d_i-D-k_i \cdot \p_i \,.
\eqe 
Now, the piece in the correlator that would survive the LSZ reduction is given as:
\eq
{\rm LSZ}\times \langle 0 |\tilde\phi_1 \cdots \tilde\phi_n | 0 \rangle={\rm LSZ}\times \frac{\delta^D(\sum_i k_i)}{\prod_{l}k_l^2}M_n+\cdots=\delta^D(\sum_i k_i)M_n\,,
\eqe
thus we find:\footnote{\label{fD}Note that the operator $\mathfrak{D}_i$ is sandwiched in between the LSZ factor and the $\frac{1}{k^2_i}$ factor form the correlator. Pushing $\mathfrak{D}_i$ past the latter acquires a factor of  $[\mathfrak{D}_i,{1\over k_i^2}]={2\over k_i^2}$.}
\eqa
\text{eq.}~(\ref{b})&=&\bigg[{1 \over 2}(d-D+1)+\sum_i(2+\mathfrak{D}_i)\bigg]\sum_j(2+\mathfrak{D}_j)\delta^D(K)M_n+\mathcal{O}(p,q),
\eqae
where $K=\sum_i k_i$.
Use the fact that $\left[\sum_i\mathfrak{D}_i,\delta^D(K)\right]=D\delta^D(K)$,
we arrive at the following double soft theorem:
\eq
M_{n+2}|_{p,q\rightarrow 0}=\left[{1 \over 2}(d-D+1)+\sum_j (2+\mathfrak{D}_j)+D\right]\left[\sum_j (2+\mathfrak{D}_j)+D\right]M_n+\mathcal{O}(p,q)\,.
\eqe
Consider the case where the remaining fields are just canonical scalars, i.e. $d_i=(D-2)/2$,\footnote{For currents, $d=1$, due to $J^\mu|0\rangle\sim p^\mu|\phi\rangle$.} the soft theorem reduces to 
\eqa
M_{n+2}|_{p,q\rightarrow 0}&=&\left[n\frac{(D-2)}{2}-D+\sum_jk_j\cdot \p_j\right]\left[(n{+1})\frac{(D-2)}{2}-D+\sum_j k_j\cdot \p_j\right]M_n\nonumber\\
&&+\mathcal{O}(p,q)\,.
\eqae
This is the leading double soft theorem due to the (broken) conformal symmetry for scalar amplitudes.
\subsubsection{Subleading double-soft theorems from $(j_{D}, j_{K}) $}
Similar to the single-soft limit, by replacing the dilatation current with that of the special conformal transformation, we can obtain further constraints for the sub-leading double-soft limit. Let us begin with following Ward identity:
\eqa
&&{\rm LSZ} \int d^Dy \, e^{ip \cdot y} {\p \over \p y^\nu} \int d^Dx \, e^{iq \cdot x} {\p \over \p x^\mu}  \langle 0 | j_D^\mu (x) j_{K^\lambda}^\nu (y) \phi(x_1) \cdots \phi (x_n) | 0 \rangle \nonumber\\
&=&{\rm LSZ} \int d^Dy \, e^{ip \cdot y}\, (-ip_\nu) \bigg[ e^{iq \cdot y} \langle 0 | \delta_D j^\nu_{K^\lambda} (y) \phi (x_1) \cdots \phi (x_n) | 0 \rangle\nonumber\\ 
&&+ \sum_i e^{iq \cdot x_i} \langle 0 | j_{K^\lambda}^\nu (y) \phi(x_1) \cdots \delta_D\phi(x_i) \cdots \phi(x_n) \bigg], \label{sub}
\eqae
where we have used the Ward identity associated with the dilatation current. Start with the first term on the RHS, and use
\eq
\delta_D\mathcal{O}(x) = (d+x\cdot\p)\mathcal{O}(x), \qquad \delta_{K^\lambda}\mathcal{O}(x) =\left[ 2x_\lambda\left(d+x\cdot \p\right)-x^2\p_\lambda\right]\mathcal{O}(x)\, ,
\eqe
we have
\eqa \label{exp}
&&{\rm LSZ} \int d^Dy \, e^{i(p+q) \cdot y} ({-}ip_\nu) \langle 0 | \delta_D j^\nu_{K^\lambda} (y) \phi (x_1) \cdots \phi (x_n) | 0 \rangle \nonumber\\
&=& \left(\prod_ik_i^2\int dx_i \, e^{ik_i\cdot x_i}\right) \left({-}ip_\nu)(d{-}D{-}(p{+}q)\cdot \p_p\right)\int dy \, e^{i(p{+}q)\cdot y}\langle 0| j_{K^\lambda}^\nu (y)\phi(x_1)\cdots \phi(x_n)|0\rangle\nonumber\\
&=&\left(\prod_ik_i^2\right)({-}ip_\nu)\left(d{-}D{-}(p{+}q)\cdot \p_p\right)\langle\tilde{j}_{K^\lambda}^\nu (p{+}q) \tilde{\phi} (k_1)\cdots \tilde{\phi}(k_n)\rangle.
\eqae
Since when acting on the vacuum $\tilde{j}_{K^\lambda}^\nu (p+q)$ excites Goldstone mode, and will generate a term proportional to ${(p+q)^\nu \over (p+q)^2}$, in the expansion of $p+q$ it can be written as
\eq
\langle\tilde{j}_{K^\lambda}^\nu (p+q) \tilde{\phi} (k_1)\cdots \tilde{\phi}(k_n)\rangle={(p+q)^\nu \over (p+q)^2}\mathfrak{S}_{-1}+\mathfrak{S}_0^\nu+(p+q)^\nu \mathfrak{S}_1+\cdots.
\eqe
Upon contraction with $(p+q)_\nu$, at leading order one must recover the Ward identity, one immediately deduce:
\eq
\mathfrak{S}_{-1}=i\sum_i\langle\tilde{\phi}_1\cdots\tilde{\delta}_{K^\lambda}\tilde{\phi}_i\cdots \tilde{\phi}_n\rangle,
\eqe
where 
\bea \tilde{\delta}_{K^\lambda}\mathcal{O}(p)=-2i\left[(d-D-p\cdot \p_p)\p_\lambda+{1\over 2}p_\lambda \p_p^2\right]\mathcal{O}(p)\equiv -2i 
\mathfrak{K}_{\lambda}\mathcal{O}(p) \, ,
\eea and we denote $\p_\lambda := \p_{p^\lambda}$.  
Thus we find:
\eqa
\text{eq.}~(\ref{exp})&=&\left(\prod_i k_i^2\right) p_\nu\left(d-D-(p+q)\cdot \p_p\right){(p+q)^\nu \over (p+q)^2}\sum_i\langle\tilde{\phi}_1\cdots\tilde{\delta}_{K^\lambda}\tilde{\phi}_i\cdots \tilde{\phi}_n\rangle +\mathcal{O}(p,q)\nonumber\\
&=&\left(\prod_i k_i^2\right){1\over 2}\left(d-D+1\right)\sum_i\langle\tilde{\phi}_1\cdots\tilde{\delta}_{K^\lambda}\tilde{\phi}_i\cdots \tilde{\phi}_n\rangle +\mathcal{O}(p,q)\nonumber\\
&=&-i \left( d-D+1\right) \sum_i \left(2 \p_{i, \lambda}+ \mathfrak{K}_{i,\lambda}\right) \delta^D(K) M_n +\mathcal{O}(p,q)\,.
\eqae
The second term of the RHS in eq.~(\ref{sub}) is a straightforward Ward identity for conformal boost, and hence we write:\footnote{Similar to footnote \ref{fD} for $\mathfrak{D}$, here pushing $\mathfrak{K}_{i, \lambda}$ past the $\frac{1}{k_i^2}$ one acquires a factor of $[\mathfrak{K}_{i, \lambda},{1\over k_i^2}]={2\p_{i, \lambda}\over k_i^2}+(D-2d_i-2){k_{i,\lambda}\over k^2}={2\p_{i, \lambda}\over k_i^2}$, where the scaling dimension of scalar $d_i={D-2 \over 2}$ has been used.}
\eqa
&&{\rm LSZ} \, \sum_{i, j} \langle 0 | \phi(x_1) \cdots \delta_{K^\lambda}\phi(x_i) \cdots \delta_D\phi(x_j) \cdots \phi(x_n) | 0 \rangle \nonumber\\
&=&-2i \sum_{i, j} k_i^2 k_j^2 \mathfrak{K}_{i,\lambda} \mathfrak{D}_{j}{1 \over k_i^2 k_j^2} \delta^D(K) M_n +\mathcal{O}(p,q)\nonumber\\
&=&-2i \sum_{i, j} \left(2 \p_{i, \lambda}+ \mathfrak{K}_{i,\lambda}\right) \left(2+ \mathfrak{D}_j\right)\delta^D(K) M_n +\mathcal{O}(p,q)\,.
\eqae
Now for the LHS of eq.~(\ref{sub}), due to the derivative relation amongst the currents, from eq.~(\ref{KDRel}) we should get
\eq
{\rm LSZ} \int d^Dy \, e^{ip \cdot y} {\p \over \p y^\nu} \int d^Dx \, e^{iq \cdot x} {\p \over \p x^\mu}  \langle 0 | j_D^\mu (x) j_{K^\lambda}^\nu (y) \phi(x_1) \cdots \phi (x_n) | 0 \rangle = -2i \p_{p, \lambda} \delta^{D}(K) M_{n+2} \, .
\eqe
Equating above results, and contracting $(p+q)^\lambda$ into both sides, we obtain the sub-leading double soft theorem:
\eqa
(p+q)\cdot \p_p\: \delta^D(K) M_{n+2} &=& (p+q)^\lambda \sum_i \left(2 \p_{i, \lambda}+ \mathfrak{K}_{i,\lambda}\right) \left[{1\over 2}\left( d-D+1\right) + \sum_j \left(2+\mathfrak{D}_j \right)\right]\nonumber\\
&&\times \delta^D(K) M_n+\mathcal{O}(p^2,p\cdot q,q^2) \, .
\eqae
Again use $\left[\sum_i\mathfrak{D}_i,\delta^D(K)\right]=D\delta^D(K)$, we get
\eqa
&&(p+q) \cdot {\p \over \p p } M_{n+2} = (p+q)^\lambda \sum_i \left(2 \p_{i, \lambda}+ \mathfrak{K}_{i,\lambda}\right) \nonumber\\
&&\times \left[{1\over 2}\left( d-D+1\right) + \sum_j \left(2+\mathfrak{D}_j\right)+D\right] M_n +\mathcal{O}(p^2,p\cdot q,q^2) \, .
\eqae
Thus express the double-soft theorem as $M_{n+2} = (S_0+(p+q)\cdot S_1)M_n$, we finally obtain the sub-leading soft factor as,
\eq
S_{1 \lambda}= \sum_i \left(2 \p_{i, \lambda}+ \mathfrak{K}_{i,\lambda}\right) \left[{1\over 2}\left( d-D+1\right) + \sum_j \left(2+\mathfrak{D}_j\right)+D\right].
\eqe
Considering the case where all other fields are scalars, i.e. $d=\frac{D-2}{2}$, we arrive at
\eq
S_{1 \lambda}= \sum_i \left[\left({D-2 \over 2}+k_i\cdot \p_i\right)\p_{i, \lambda}-{1\over 2}k_{i,\lambda} \p_i^2\right] \left[(n+1){D-2 \over 2}-D+\sum_j k_j\cdot \p_j\right] \, ,
\eqe 
which is the sub-leading double soft theorem of dilatons when scatter with scalars.

\subsection{Broken Translational symmetry}
Here we consider another kind of spontaneous broken space-time symmetry, broken translation and Lorentz rotation due to the presence of branes. The low energy effective actions for the brane describes the interaction of the Goldstone bosons associated with broken translation and Lorentz rotation. We consider the case where a $D$-dimensional brane is embedded in $(D{+}n)$-dimensional flat space, with $n=1$. It is straight forward to extend it to general $n$. The space time index is separated as $x^M=\{x^\mu,x^0\}$, where $\mu$ denote the longitudinal directions along the brane and $0$ denotes the transverse direction.   

In the presence of a brane, both $P_0$ and $L_{0\mu}$ are broken. However, as the case of (broken) conformal symmetry, there is only a single Goldstone mode. That is because $P_0$ and $L_{0\mu}$ are related by
\eq
[L_{0\mu}, P_{\nu}]=\eta_{\mu\nu}P_0 \,,
\eqe  
the mode associated with broken translation symmetry is derivatively related to that of broken Lorentz transformation. More precisely, since the current must be conserved, on dimension grounds one can deduce 
\eq\label{PLRel}
\p_\mu j^\mu_{L_{0\nu}}\sim x^\nu\p_\mu j^\mu_{P_{0}}\,,
\eqe 
similar to the case for broken conformal symmetries. Thus we would expect universal leading and sub-leading soft theorems, as we will show in the following. 
\subsubsection{Single soft theorems}
We begin by considering the following LSZ reduction of the current correlator,
\eq\label{SingleSoftP} 
{\rm LSZ}\int d^Dx\,e^{ip\cdot x}\,\frac{\p}{\p x^\mu}\langle  j^\mu_{P_{0}}(x) j^{\mu_1}_{P_{0}}(x_1)\cdots j^{\mu_n}_{P_{0}}(x_n)\rangle \, .
\eqe
Apply the Ward identity and the fact that $\delta_{P_0}j^\mu_{P_{0}}\sim [j^\nu_{P_{0}},j^\mu_{P_{0}}]=0$, one finds that eq.~(\ref{SingleSoftP}) vanishes. On the other hand using eq.~(\ref{ward}) and eq.~(\ref{SingleSoftP}) yields
\eq
p_\mu\left(\prod_{i=1}^n k^{2}_i\right)\langle  \tilde{j}^\mu_{P_{0}}(p) \tilde{j}^{\mu_1}_{P_{0}}(k_1)\cdots \tilde{j}^{\mu_n}_{P_{0}}(k_n)\rangle=M_{n+1}(\pi(p)\pi(k_1)\cdots \pi(k_n))\left( \prod_{i=1}^n k^{\mu_i}_i\right)+\mathcal{O}(p).
\eqe
So we conclude that, 
\eq \label{eq:P0}
\left.M_{n+1}(\pi(p) \pi(k_1)\cdots \pi(k_n))\right|_{p\rightarrow 0}=0+\mathcal{O}(p) \, .
\eqe
This is the Adler's zero for broken translation symmetry. 

Now, instead use $j^\mu_{L_{0\lambda}}$ and eq.~(\ref{PLRel}), we have the relation,
\eq\label{Temp} 
\int d^Dx\,e^{ip\cdot x}\,\frac{\p}{\p x^\mu}\langle j^\mu_{L_{0\lambda}}(x) j^{\mu_1}_{P_{0}}(x_1)\cdots j^{\mu_n}_{P_{0}}(x_n)\rangle=\int d^Dx\,e^{ip\cdot x}\,x^\lambda\frac{\p}{\p x^\mu}\langle j^\mu_{P_{0}}(x) j^{\mu_1}_{P_{0}}(x_1)\cdots j^{\mu_n}_{P_{0}}(x_n)\rangle\,.
\eqe
Once again apply the Ward identity, since $\delta_{L_{0\lambda}}j^\mu_{P_{0}}\sim [j^\mu_{L_{0\lambda}},j^\mu_{P_{0}}]=j^\mu_{P_{\lambda}}$, thus it does not excite a physical state, one finds:
\eq\label{SingleSoftL} 
{\rm LSZ}\int d^Dx\,e^{ip\cdot x}\,\frac{\p}{\p x^\mu}\langle j^\mu_{L_{0\lambda}}(x) j^{\mu_1}_{P_{0}}(x_1)\cdots j^{\mu_n}_{P_{0}}(x_n)\rangle=0.
\eqe 
On the other hand, using the fact that $j^\mu_{P_{0}}(x)$ excites a Goldstone mode, we can also write eq.~(\ref{Temp}) as:
\eqa
&&\left(\prod_{i=1}^n k^{2}_i\right)\frac{\p}{\p p^\lambda} p_\mu \langle \tilde{j}^\mu_{P_{0}}(p) \tilde{j}^{\mu_1}_{P_{0}}(k_1)\cdots \tilde{j}^{\mu_n}_{P_{0}}(k_n)\rangle=\nonumber\\
&& \frac{\p}{\p p^\lambda}\left[ M_{n{+}1}(\pi(p)\pi(k_1)\cdots \pi(k_n))\prod_{i=1}^n\left( k^{\mu_i}_i\right)+\mathcal{O}(p)\right].
\eqae
Using the fact that $M_{n+1}$ vanishes in the soft limit from eq.~(\ref{eq:P0}), therefore expanding in terms of the soft momentum we have $M_{n+1}=\sum_{a=1}p^a m_a$. Then the above equation implies that the leading term $m_1=0$, and hence 
\eq
\left.M_{n{+}1}(\pi(p)\pi(k_1)\cdots \pi(k_n))\right|_{p\rightarrow 0}=0+\mathcal{O}(p^2)\,,
\eqe
i.e. spontaneously broken translational symmetries implies that amplitudes involving a soft Goldstone boson vanishes up to order $\mathcal{O}(p^2)$ in the soft momentum.

\subsubsection{Leading double soft from $(j_{P_{0}}, j_{P_0}) $}
Just as the case of the single-soft theorems we just discussed, one can apply Ward identity with $j_{P_0}$ twice to obtain the leading double-soft theorem. To do so, we start with following identity,  
\eqa
&&{\rm LSZ}\int d^Dy\, e^{iq\cdot y}\frac{\p}{\p y^\nu}\int d^Dx\, e^{ip\cdot x} \frac{\p}{\p x^\mu} \langle 0 | j_{P_0}^\mu (x)j_{P_0}^\nu (y) j_{P_0}^{\mu_1} (x_1) \cdots j_{P_0}^{\mu_n} (x_n) | 0 \rangle\nonumber\\
&=&{\rm LSZ}\int d^Dy\, (-iq^\nu)\bigg\{e^{i(q+p)\cdot y} \langle 0 | \delta_{P_0} j_{P_0}^\nu (y) j_{P_0}^{\mu_1} (x_1) \cdots j_{P_0}^{\mu_n} (x_n) | 0 \rangle\nonumber\\
&&+\sum_{i=1}^ne^{i(p\cdot x_i+q\cdot y)}\langle 0 |  j_{P_0}^\nu (y) j_{P_0}^{\mu_1} (x_1)\cdots \delta_{P_0}j_{P_0}^{\mu_i}(x_i) \cdots j_{P_0}^{\mu_n} (x_n) | 0 \rangle\bigg\}=0\,,
\eqae
where we have used that $\delta_{P_0}j_{P_0}^{\mu_i}=0$. On the other hand we also have: 
\eqa
&&{\rm LSZ}\int d^Dy\, e^{iq\cdot y}\frac{\p}{\p y^\nu}\int d^Dx\, e^{ip\cdot x} \frac{\p}{\p x^\mu} \langle 0 | j_{P_0}^\mu (x)j_{P_0}^\nu (y) j_{P_0}^{\mu_1} (x_1) \cdots j_{P_0}^{\mu_n} (x_n) | 0 \rangle\nonumber\\
&=&\left( \prod_i k_i^2\right)q_\nu p_\mu \langle 0 | \tilde{j}_{P_0}^\mu (q)\tilde{j}_{P_0}^\nu (p) j_{P_0}^{\mu_1} (x_1) \cdots j_{P_0}^{\mu_n} (x_n) | 0 \rangle\nonumber\\
&=& M_{n+2}(\pi(q)\pi(p)\pi(k_1)\cdots \pi(k_n))\prod_{i=1}^n\left( k^{\mu_i}_i\right)+\mathcal{O}(p,q)\,.
\eqae
Thus combining these two results we conclude that the double-soft limit vanishes up to leading order, i.e. \bea
M_{n+2}(\pi(q)\pi(p)\pi(k_1)\cdots \pi(k_n))=0 + \mathcal{O}(p,q) \, .
\eea
\subsubsection{Subleading double-soft theorems from $(j_{L_{0\lambda}}, j_{P_0}) $}
To obtain higher-order double soft theorems, we instead consider correlators involving both $j_{L_{0\lambda}}$ and $j_{P_0}$. In particular, we will study, 
\eqa\label{DL}
&&{\rm LSZ}\int d^Dy\, e^{iq\cdot y}\frac{\p}{\p y^\nu}\int d^Dx\, e^{ip\cdot x} \frac{\p}{\p x^\mu} \langle 0 | j_{L_{0\lambda}}^\mu (x)j_{P_0}^\nu (y) j_{P_0}^{\mu_1} (x_1) \cdots j_{P_0}^{\mu_n} (x_n) | 0 \rangle\nonumber\\
&=&{\rm LSZ}\int d^Dy\, (-iq_\nu)\Big\{e^{i(q+p)\cdot y} \langle 0 | \delta_{L_{0\lambda}} j_{P_0}^\nu (y) j_{P_0}^{\mu_1} (x_1) \cdots j_{P_0}^{\mu_n} (x_n) | 0 \rangle\nonumber\\
&&\left.+\sum_{i=1}^ne^{i(p\cdot x_i+q\cdot y)}\langle 0 |  j_{P_0}^\nu (y) j_{P_0}^{\mu_1} (x_1)\cdots \delta_{L_{0\lambda}}j_{P_0}^{\mu_i}(x_i) \cdots j_{P_0}^{\mu_n} (x_n) | 0 \rangle\right\}\,.
\eqae
Since $\delta_{L_{0\lambda}}j_{P_0}\sim[j_{L_{0\lambda}},j_{P_0}]=j_{P^\lambda}$, the last line in the above does not survive the LSZ reduction. The contribution of the first line is given as:
\eq
(-iq_\nu)\sum_{i=1}^n\left[\langle \pi(k_i)| \widetilde{\delta_{L_{0\lambda}} j_{P_0}^\nu}(p{+}q)|\phi(k_i{+}q{+}p)\rangle\frac{1}{2k_i\cdot(p{+}q)}\langle\phi(k_i{+}q{+}p)\cdots \rangle\right]\left(\prod_{i=1}k_i^{\mu_i}\right) \,.
\eqe
We will evaluate $\langle \pi(k_i)| \widetilde{\delta_{L_{0\lambda}} j_{P_0}^\nu}(p{+}q)|\phi(k_i{+}q{+}p)\rangle$ by considering the following diagrammatic representation of this term, 
\eq
\includegraphics[scale=0.5]{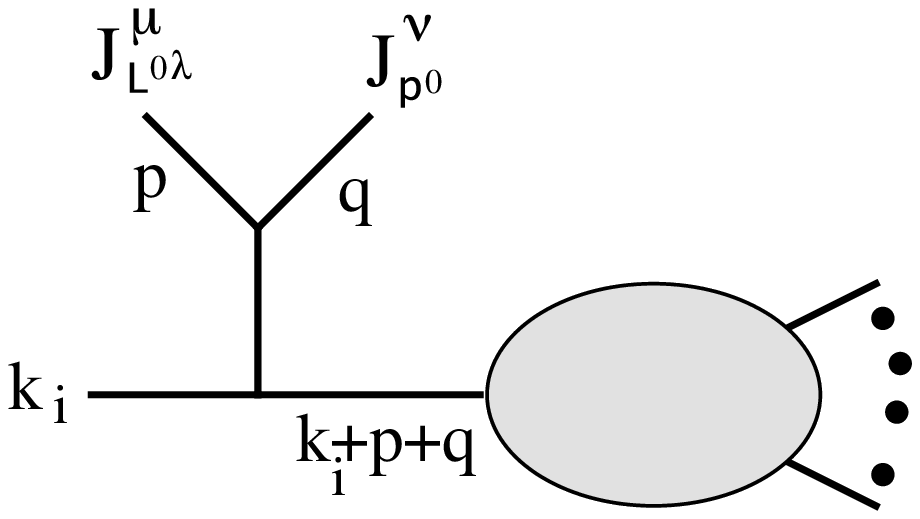}
\eqe
Since $[L^{0\lambda},P^0]=P^\lambda$, the contribution of this diagram should be given by,
\eqa
&& \alpha \frac{(p+q)_\lambda}{(p+q)^2}\frac{p_\mu (2k_i+p+q)^\mu q_\nu (2k_i+p+q)^\nu}{2k_i\cdot(p{+}q)}M_n=\alpha  \frac{(p+q)_\lambda}{(p+q)^2}\frac{2(p_\cdot k_i) (q_\cdot k_i) }{k_i\cdot(p{+}q)}M_n+\mathcal{O}(p,q)\nonumber\\
\eqae
where $\alpha$ is an undetermined coefficient so far. Summing over all $k_i$, the RHS of eq.~(\ref{DL}) up to $\mathcal{O}(p,q)$ can be recast as,
\eq
\alpha\sum_{i}  \frac{(p+q)_\lambda}{(p+q)^2}\frac{2(p_\cdot k_i) (q_\cdot k_i) }{k_i\cdot(p{+}q)}M_n=\alpha\sum_{i}  \frac{(p+q)_\lambda}{(p+q)^2}\frac{((p-q)\cdot k_i)^2 }{2k_i\cdot(p{+}q)}M_n+\mathcal{O}(p,q)\,.
\eqe
On the other hand, using eq.~(\ref{PLRel}), the double soft limit of eq.~(\ref{DL}) can also be written as 
\eqa
&&-i\prod_{i=1}^n\left[ k^{2}_i\right]\p_{p^\lambda}q_\nu p_\mu \langle 0 | \tilde j_{P_{0}}^\mu (p)\tilde j_{P_0}^\nu (q) \tilde j_{P_0}^{\rho_1} (k_1) \cdots \tilde j_{P_0}^{\rho_n} (k_n) | 0 \rangle|_{p,q\rightarrow0}\nonumber\nonumber\\
&=&-\left.i\p_{p^\lambda}M_{n+2}(\pi(q)\pi(p)\pi(k_1)\cdots \pi(k_n))\prod_{i=1}^n\left[ k^{\nu_i}_i\right]\right|_{p,q\rightarrow0}+\mathcal{O}(p^0,q^0)\,.
\eqae
Equating these two distinct representations, and contracting with $(p+q)$, one finds:
\eq\label{Final}
-i\left.(p+q)\cdot \p_{p}M_{n+2}(\pi(q)\pi(p)\pi(k_1)\cdots \pi(k_n))\right|_{p,q\rightarrow0}=i\alpha\sum_{i} \frac{((p-q)\cdot k_i)^2 }{2k_i\cdot(p{+}q)}M_n\,.
\eqe
The undetermined constant $\alpha$ can be fixed by checking the above formula eq.~(\ref{Final}) with explicit simple tree-level amplitudes in the theory. For instance for $n=4$, namely the double-soft limit which take the six-point amplitude to the four-point one, we find $\alpha= {1 \over 2}$. We can now also compare this to the result in~\cite{Cachazo:2015ksa}, where at the leading non-vanishing order, 
\eq
\left.M_{n+2}(\pi(q)\pi(p)\pi(k_1)\cdots \pi(k_n))\right|_{p,q\rightarrow0}=\sum_i S^{(0)}_i M_{n}(\pi(k_1)\cdots \pi(k_n))+\mathcal{O}(p^2,q^2)\,,
\eqe
with the soft factor $S^{(0)}_i = \frac{1}{4}\frac{ (k_i\cdot (p-q))^2}{k_i\cdot(p+q)}$. 
Acting with $(p+q)\cdot \p_{p}$ on the above one indeed finds eq.~(\ref{Final}) with $\alpha=\frac{1}{2}$! Thus we conclude that the sub-leading soft theorem derived in~\cite{Cachazo:2015ksa} has a current algebra origin. A similar analysis using $(L^{0\mu}, L^{0\nu})$ should lead to double-soft theorem at one further higher order~\cite{Cachazo:2015ksa}, 
\bea \label{eq:S0S1}
\left.M_{n+2}(\pi(q)\pi(p)\pi(k_1)\cdots \pi(k_n))\right|_{p,q\rightarrow0}=\sum_i (S^{(0)}_i + S^{(1)}_i) M_{n}(\pi(k_1)\cdots \pi(k_n)) \,, 
\eea
where the higher-order soft factor $S^{(1)}_i$ is given by
\bea
S^{(1)}_i =\frac{1}{2} \left( -\frac{ (k_i\cdot p)^2 + (k_i\cdot q)^2}{ ( k_i\cdot(p+q))^2 } (p \cdot q) 
+
{ k_i \cdot(p-q) \over k_i \cdot(p+q) }  (p_{\mu} q_{\nu} J_i^{\mu \nu}) 
\right)\, ,
\eea
with the angular momentum $J_i^{\mu \nu}$ for scalars defined by, 
\bea
J_i^{\mu \nu} = k_i^{\mu} {\partial \over \partial {k_{i, \nu }}} -  k_i^{\nu} {\partial \over \partial {k_{i, \mu }}} \, .
\eea


\section{UV-divergence} \label{sec:UV-divergence}

\subsection{UV-divergence of DBI}

As we mentioned it is now well-known that the soft theorems we derived in previous sections are satisfied for the tree-level S-matrix in various effective field theories. Here we will study the fate of the soft theorems against the UV divergences. As examples we will consider the UV divergences of one-loop amplitudes in DBI, conformal DBI in $D=4$ and $D=6$, as well as the A-V theory for Goldstinos in $D=4$. We will verify by explicit loop computations that all the soft theorems that are derivable from current algebra as we have shown in previous sections should be respected even in the presence of UV divergences. 

\begin{figure}
\begin{center}
\includegraphics[scale=0.6]{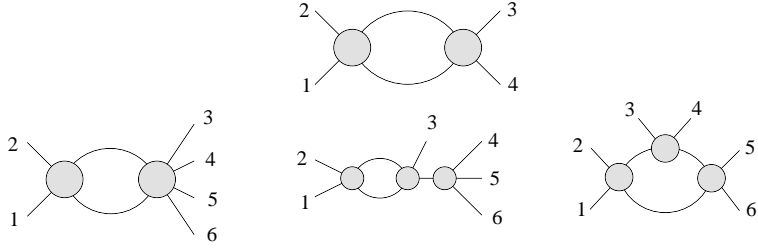}
\caption{The Feynman diagrams contribute to four and six-point amplitudes in DBI at one-loop order. One should also sum over all other independent permutations. }
\label{fig:six-pt}
\end{center}
\end{figure}

 
We begin with the DBI action of a single scalar, which takes the following form, 
\bea
\mathcal{L}_{\rm DBI} = g^{-2} \left( \sqrt{ - {\rm det} ( \eta_{\mu \nu} - g^{2} \partial_{\mu} \phi\partial_{\nu} \phi   )} - 1 \right) \, ,
\eea
where $g$ is the dimensionful coupling constant. 
Expand the square root to the order relevant for the four- and six-point amplitudes, we have, 
\bea
\mathcal{L}_{\rm DBI} = - {1 \over 2} ( \partial \phi )^2 - 
{g^2 \over 2!} \left( { ( \partial \phi )^2 \over 2 }\right)^2 - {3 g^4 \over 3!} 
\left( { ( \partial \phi )^2 \over 2 }\right)^3  + \ldots \, . 
\eea
From the action, it is then straightforward to compute the amplitudes from Feynman diagrams. In particular, the interacting vertices at four and six points are given by,  
\bea
V_4(k_1, k_2, k_3, k_4) &=& {g^2 } \left( k_1 \cdot k_2 \, k_3 \cdot k_4 + k_1 \cdot k_3 \, k_2 \cdot k_4
+ k_1 \cdot k_4 \, k_2 \cdot k_3  \right)\,,  \nonumber \\
V_6(k_1, k_2, k_3, k_4,k_5,k_6) &=& { 3 g^4  } \left( k_1 \cdot k_2 \, k_3 \cdot k_4 \, k_5 \cdot k_6 +
 \ldots \right) \, ,
\eea 
where the ellipsis in the six-point vertex $V_6$ denotes all other $14$ independent contractions for the six-point momenta. 
We then glue the above vertices to form one-loop diagrams, the Feynman diagrams that contribute to four and six-point one-loop amplitudes in DBI are shown in Fig. \ref{fig:six-pt}. In the following subsections we will study four and six-point one-loop amplitudes respectively, including both the $D=4$ and $D=6$ case.

\subsubsection{Four points at $D=4$ and $D=6$}

At four points, the one-loop integrand is obtained by gluing two four-point vertices $V_4$, which is given by the bubble diagram at the top of Fig. \ref{fig:six-pt}. Explicitly, the integrand takes the following form, 
\bea
I_4 = {1 \over 4} { \mathcal{N}_4 \over  \ell^2 (\ell+ k_1 + k_2)^2 } \, ,
\eea
where the factor ${1 \over 4}$ is the symmetry factor of the diagram, and the numerator $\mathcal{N}_4$ is defined as
\bea \label{eq:N4}
\mathcal{N}_4 =  V_4(k_1, k_2, \ell, \ell')  V_4( k_3, k_4, -\ell, -\ell')  \, ,
\eea
with $\ell' = -(\ell + k_1 + k_2)$. It is straightforward to perform the one-loop integration of the bubble integral. To be concrete we use dimensional regularization, and we are interested in the UV divergent part. The integrated result clearly depends on the space-time dimensions. Here we consider $D=4$ and $D=6$ as interesting examples. At $D=4$, we find the UV divergent part is given by\footnote{Here and in the following, we will simply ignore the coupling $g$ dependence since it is not crucial for our discussion. }
\bea \label{eq:4pt4D}
A_{4, \rm UV}^{(D=4)} = { 7 \pi^2 \over 5\epsilon } (s^4 + t^4 + u^4) \, .
\eea
In the language of UV counter terms, it shows that the UV counter term of $4D$ DBI action takes the form of $-{ 7 \pi^2 \over 5 } (\partial^8 \phi^4)$, where the matrix element of $(\partial^8 \phi^4)$ is given by $(s^4 + t^4 + u^4)$. Similarly, at $D=6$, the UV divergent part of the one-loop four-point amplitude is given 
\bea \label{eq:4pt6D}
A_{4, \rm UV}^{(D=6)} = -{ 223 \pi^3 \over 525 \epsilon} (s^5 + t^5 + u^5) \, .
\eea
So again the UV counter term may be written as ${ 223 \pi^3 \over 525 \epsilon } (\partial^{10} \phi^4)$ now for $D=6$. 

\subsubsection{Six points at $D=4$ and $D=6$}

We then consider the six-point amplitudes. There are three types of Feynman diagrams at six points as shown in Fig. \ref{fig:six-pt}. To express the results in a compact form, it is convenient to expand all the answers in terms of polynomial basis. At six points, it is easy to see that the UV divergence goes as $s^5$ in $D=4$. At this order and $D=4$, there are five independent local polynomials, and one term with a factorization pole, 
\bea 
b^{(5)}_{1} &=& s^5_{12} + \mathcal{P}_6 \, , \quad
b^{(5)}_{2} = s^5_{123} + \mathcal{P}_6 \, , \quad
b^{(5)}_{3} = s^3_{123}s_{45}^2 + \mathcal{P}_6 \, , \quad 
b^{(5)}_{4} = s^2_{123}s_{45}^3 + \mathcal{P}_6 \, , \nonumber   \\   
b^{(5)}_{5} &=& s^2_{12}s_{34}^3 + \mathcal{P}_6  \, , \quad 
F^{(5)}_{1} = {  (s^2_{12}+s^2_{23}+s^2_{13})(s^4_{45}+s^4_{46}+s^4_{56})  \over s_{123} }+ \mathcal{P}_6 \,,
\label{eq:basiss5}
\eea
where $\mathcal{P}_6$ means that we sum over full permutations on the six-point external legs.  
For convenience, we will denote this length-six list of basis as $B^{(5)}_{6}:=\{b^{(5)}_{1}, b^{(5)}_{2}, b^{(5)}_{3}, b^{(5)}_{4}, b^{(5)}_{5}, F^{(5)}_{1} \}$. 

Let us now discuss each contribution to the six-point amplitude in Fig. \ref{fig:six-pt}. Begin with the first bubble diagram in Fig. \ref{fig:six-pt}, the integrand is given by
\bea
I^{(1)}_6 = {1 \over 2} {\mathcal{N}^{(1)}_6 \over \ell^2 (\ell + k_1 + k_2)^2 } \,, 
\eea
and the numerator is the product of a four-point vertex and a six-point vertex as shown in the figure,
\bea
\mathcal{N}^{(1)}_6 = V_4(k_1, k_2, \ell, \ell') V_6(k_3, k_4, k_5, k_6, -\ell, -\ell')
 \, ,
\eea
where $\ell' = -(\ell + k_1 + k_2)$. Performing the integration, we find the ten-derivative UV divergence. Expressed in terms of the basis $B^{(5)}_{6}$ defined previously, the contribution of this particular diagram is given by $C^{(D=4)}_1  \cdot B^{(5)}_{6}$, where we find that the coefficient $C^{(D=4)}_1 $ is given by, 
\bea
\epsilon C^{(D=4)}_1 = \biggr\{ {181 \pi^2 \over 3600},  - {2 \pi^2 \over 675}, -{8 \pi^2 \over 45}, - {\pi^2 \over 180}, {11 \pi^2 \over 40},0 \biggr\} \, .
\eea
The integrand of the second bubble diagram in Fig. \ref{fig:six-pt} with a factorization pole takes the following form, 
\bea
I^{(2)}_6 = {1 \over 2} {\mathcal{N}^{(2)}_6 \over \ell^2 (\ell + k_1 + k_2)^2 s_{456} } \,, 
\eea
and the numerator now is the product of three four-point vertices, 
\bea
\mathcal{N}^{(2)}_6 = V_4(k_1, k_2, \ell, \ell')V_4(k_3, P, -\ell, -\ell') V_4 (k_4, k_5, k_6, -P) \,,
\eea
with $\ell' = -(\ell + k_1 + k_2)$, and $P= k_4+k_5+k_6$. The UV divergence of this diagram is given by $C^{(D=4)}_2 \cdot B^{(5)}_{6}$, with the coefficient,
\bea
\epsilon C^{(D=4)}_2= \biggr\{ -{ 77 \pi^2  \over 2700},  {13 \pi^2 \over 16200}, {13 \pi^2 \over 270}, {\pi^2 \over 360}, -{59 \pi^2 \over 360}, {7 \pi^2 \over 360} \biggr\} \, .
\eea
Finally, the triangle diagram with three four-point vertices in Fig.~\ref{fig:six-pt} has the integrand of the form
\bea
I^{(3)}_6 = {1 \over 3!} {\mathcal{N}^{(3)}_6 \over \ell^2 (\ell + k_1 + k_2)^2 (\ell - k_3 - k_4)^2} \,, 
\eea
and the numerator is given by
\bea
\mathcal{N}^{(3)}_6 =  V_4(k_1, k_2, \ell, \ell_1) V_4(k_3, k_4, -\ell, \ell_2)  V_4(k_5, k_6, -\ell_1, -\ell_2)  \,,
\eea
where $\ell_1 = -(\ell + k_1 + k_2)$ and  $\ell_2 = \ell - k_3 - k_4$. Perform the integration, we find the UV divergence of this diagram is given by $C^{(D=4)}_3 \cdot B^{(5)}_{6}$, with 
\bea
\epsilon C^{(D=4)}_3 =\biggr\{ -{ 319 \pi^2 \over 10800}, -{91  \pi^2 \over 16200}, {7 \pi^2 \over 54}, { \pi^2 \over 45}, -{41 \pi^2 \over 180}, 0\biggr\}
\eea
Put all the contributions from three diagrams together, we have
the full one-loop UV divergent part at six points of DBI action at $D=4$, 
\bea
A_{6, \rm UV}^{(D=4)} = C^{(D=4)}_1 \cdot B^{(5)}_{6} +  C^{(D=4)}_2 \cdot B^{(5)}_{6}  +   C^{(D=4)}_3 \cdot B^{(5)}_{6} \, .
 \eea
This finishes the computation of the UV divergence of the six-point amplitude in $4D$ DBI action, and we will discuss its soft limits shortly.    

We now consider the six-point DBI amplitude at $D=6$ as another example. The Feynman diagrams and the loop integrands are of course independent of the space-time dimensions.  The power counting of the six-point UV divergence is now of order $s^6$ for $D=6$. Again we will express the results in terms of polynomial basis, which has $13$ independent local polynomials at this order. Here is the list of the independent basis elements as well as one term with a factorization pole:  
\bea
b_1^{(6)} &=& s_{12}^6 + \mathcal{P}_6 \, , \quad
b_2^{(6)} = s_{123}^6 + \mathcal{P}_6 \, , \quad 
b_3^{(6)} = s_{12}^4s_{13}^2 + \mathcal{P}_6  \, , \\ \nonumber
b_4^{(6)} &=& s_{12}^4 s_{34}^2 + \mathcal{P}_6 \, , \quad
b_5^{(6)} = s_{12}^3 s_{13}^3+ \mathcal{P}_6 \, , \quad 
b_6^{(6)} = s_{12}^3 s_{34}^3 + \mathcal{P}_6 \, , \\ \nonumber
b_7^{(6)} &=& s_{12}^2 s_{123}^4+ \mathcal{P}_6 \, , \quad 
b_8^{(6)} = s_{14}^2 s_{123}^4+ \mathcal{P}_6 \, ,\quad
b_9^{(6)} = s_{14}^4 s_{123}^2 + \mathcal{P}_6 \, , \\ \nonumber
b_{10}^{(6)} &=& s_{14}^3 s_{123}^3  + \mathcal{P}_6 \, ,\quad 
b_{11}^{(6)} = s_{123}^3 s_{124}^3  + \mathcal{P}_6 \, ,\quad 
b_{12}^{(6)} =s_{12}^2 s_{34}^2 s_{56}^2 + \mathcal{P}_6 \, , \\ \nonumber
b_{13}^{(6)} &=& s_{123}^2 s_{124}^2 s_{135}^2 + \mathcal{P}_6 \, , \quad 
F_{1}^{(6)} = { (s_{12}^5+s_{23}^5+s_{13}^5) (s_{45}^2+s_{56}^2+s_{61}^2)  \over s_{123} }  + \mathcal{P}_6 \, .
\eea
These independent basis form a list, which we will denote as $B^{(6)}_{6}$. The first bubble diagram with four and six-point vertices in terms of the coefficient of these $14$ basis is given by
\bea
 \epsilon{C^{(D=6)}_1 } &=&   \biggr\{ -{181 \pi^3 \over 11200 }, -{\pi^3 \over 4200}, -{17 \pi^3 \over 840}, -{33\pi^3 \over 560}, -{ \pi^3 \over 70}, 
-{79 \pi^3 \over  3360}, {23 \pi^3\over 3360},
\cr  && - {11 \pi^3 \over 2240}, {199 \pi^3 \over 3360}, {\pi^3 \over 1680}, 
-{11 \pi^3 \over 3360}, {33 \pi^3 \over 2240}, 0, 0 \biggr\} \, .  
\eea
While for the second bubble diagram, the coefficient is now 
\bea
\epsilon C^{(D=6)}_2 &=& \biggr\{ { 2809 \pi^3 \over 302400}, {\pi^3 \over 12600},  {53 \pi^3 \over 7560}, {431 \pi^3 \over 10080}, {\pi^3 \over 210}, {121 \pi^3 \over 15120}, 
 -{43 \pi^3 \over 18900}, {\pi^3 \over 630},  -{ 89 \pi^3 \over 5040}, 
 \cr
 && -{\pi^3 \over 7560}, {17 \pi^3 \over 15120}, -{17 \pi^3 \over 3360}, 0, -{223  \pi^3 \over 37800}
  \biggr\}
\eea
Finally the coefficient for the triangle diagram with three four-point vertices is
\bea
\epsilon C^{(D=6)}_3 &=& \biggr\{{1823 \pi^3 \over 362880}, {5 \pi^3 \over 13608}, -{187 \pi^3 \over 22680}, 
{3601 \pi^3 \over 60480}, -{19 \pi^3 \over 1890}, -{799 \pi^3 \over 90720}, 
{  79 \pi^3 \over 45360}, -{83 \pi^3 \over 12096},
\cr
&& -{727 \pi^3 \over 30240}, {337 \pi^3 \over 45360}, -{779 \pi^3\over 90720}, {8 \pi^3\over 105}, 0, 0 \biggr\}\, .
\eea
The final result is the sum of these three contributions, 
\bea
A^{(D=6)}_{6, \rm UV}= C^{(D=6)}_1 \cdot B^{(6)}_{6} + C^{(D=6)}_2 \cdot B^{(6)}_{6}  +  C^{(D=6)}_3 \cdot B^{(6)}_{6} \, .
 \eea
With these explicit one-loop results, we have checked that both $A_{\rm UV}^{(D=4)}$ and $A_{\rm UV}^{(D=6)}$ indeed satisfy the single and double-soft theorems, namely, 
\bea
A_{6, \rm UV}^{(D)}\big{|}_{k_6 \rightarrow \tau k_6} \sim \mathcal{O}(\tau^2) \, ,
\eea
and the double-soft theorem, 
\bea
A_{6, \rm UV}^{(D)}\big{|}_{k_5 \rightarrow \tau k_5, \,  k_6 \rightarrow \tau k_6} =  
\left( \tau S^{(0)}  + \tau^2 S^{(1)} \right)  A_{4, \rm UV}^{(D)} \, ,
\eea
where $S^{(0)}$ and $S^{(1)}$ are the  leading and subleading double-soft factors defined in eq.~(\ref{eq:S0S1}). 

The six-point UV divergence $A_{6, \rm UV}^{(6)}$ in fact further satisfies the order $\mathcal{O}(\tau^3)$ double-soft theorem of~\cite{Cachazo:2015ksa}, namely it has the same soft behaviour as the tree-level DBI amplitudes. This result can be easily understood once one realises that the original order $\mathcal{O}(\tau^3)$ double-soft theorem is a consequence of the four-point structure being of the form $s^2+t^2+u^2$. In the double soft limit, the relevant diagram is of the following form:
$$\includegraphics[scale=0.5]{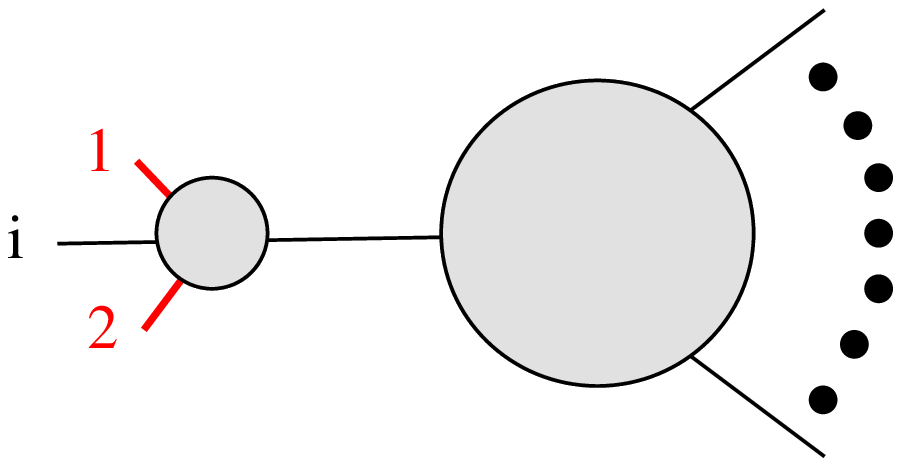}$$
Since the original four-point vertex is of four derivatives, the leading contribution from the four-vertex in the soft limit, $k_1 \rightarrow \tau k_1$ and $k_2 \rightarrow \tau k_2$ for $\tau \rightarrow 0$, would be of the form $\tau{ (k_1 \cdot k_i)^2 + (k_2 \cdot k_i)^2 \over (k_1 + k_2)\cdot k_i }$ which indicates the double soft-limit begins at $\mathcal{O}(\tau)$. Now the UV-divergence introduces a new vertex  $s^4+t^4+u^4$, which in the soft limit takes the form  
\bea
\left.{(s_{12}^4 + s_{1 i}^4 + s_{2 i}^4) \over s_{12i} } {\times} A_R \right|_{k_1 \rightarrow \tau k_1, k_2 \rightarrow \tau k_2} = 
\tau^3 {8( (k_1 \cdot k_i)^4 + (k_2 \cdot k_i)^4) \over (k_1 + k_2)\cdot k_i } {\times} A_R + 
\mathcal{O}(\tau^4) \,, 
\eea
and thus one expects the order $\mathcal{O}(\tau^3)$ double-soft theorem of~\cite{Cachazo:2015ksa} to be modified. However, at $D=6$ the four-point UV divergence goes as $s^5+t^5+u^5$, as shown in eq.~(\ref{eq:4pt6D}), which would only contribute to the order $\mathcal{O}(\tau^4)$ and thus leave the double-soft theorem untouched. We have further checked the UV divergence at $D=8$ also satisfies the double-soft theorem to the order $\mathcal{O}(\tau^3)$, since just by power counting its four-point divergence goes as even higher order, namely $s^6$.

It is instructive to understand the above result from recursion. As discussed in~\cite{Cheung:2015ota}, the single-soft theorem at order $\mathcal{O}(\tau)$ ensures that the $(2n)$-point amplitudes of order $s^m$ with $m<2n$ are soft on-shell constructible. This is indeed the case for tree-level DBI amplitudes, where at $2n$-points it behaves as $s^{n}$. The $n$-point one-loop UV divergence for DBI at $D$-dimensions goes as $s^{(n+D)/2}$, which means that if single soft theorems are respected, then for $D=4$ all the higher-point UV divergent terms are completely fixed by the four-point. On the other hand since the single-soft theorems are symmetry based, this implies that if the $D=4$ UV divergent violates any tree-level soft theorems, the later cannot hold solely on symmetry grounds.  

For $D=6$, the one-loop UV divergence are determined by the four and six-point ones which we have computed explicitly in this paper. A similar conclusion regarding how soft theorems are able to constrain the results of UV counter terms can also apply to other theories such as the conformal DBI and the Akulov-Volkov theory as well as the $\alpha'$ corrections from string theories, which we will discuss in the next sections. 

\subsection{Conformal DBI} 

To test the soft theorems of conformal symmetry, we will study the one-loop UV divergence of conformal DBI.\footnote{The leading and subleading single-soft theorems due to (broken) conformal symmetry have also been tested in detail in~\cite{Bianchi:2016viy} for the lower-energy effective action of $\mathcal{N} = 4$ super Yang-Mills (SYM) on the Coulomb branch, both perturbatively at one-loop order and
non-perturbatively via the one-instanton effective action~\cite{Bianchi:2015cta}. } The conformal DBI action (with single scalar) takes the form, 
\bea
\mathcal{S}_{\rm CDBI} &=& \int d^D x \, {\phi^D} \left( \sqrt{ - {\rm det} ( \eta_{\mu \nu} - {1 \over \phi^4} \partial_{\mu} \phi\partial_{\nu} \phi   )} - 1 \right) \cr
&=& 
\int d^D x \, { \phi^D} \left( - {1 \over 2 \phi^4} \partial \phi \cdot \partial \phi
- {1 \over 8 \phi^8} (\partial \phi \cdot \partial \phi)^2  + \ldots \right)   \, ,
\eea
here we only expand to the order which is relevant to the computation in this section. 
We will consider the theory in $D=4$ as well as $D=6$, as the case of flat space DBI. Therefore, for $D=4$ we have
\bea
\mathcal{S}_{\rm CDBI} 
=
\int d^4 x \, \left( - {1 \over 2 } \partial \phi \cdot \partial \phi
- {1 \over 8 \phi^4} (\partial \phi \cdot \partial \phi)^2  + \ldots \right)   \, ,
\eea
The action should be understood with $\phi = v + \phi$, where $v$ is the vev which breaks the conformal symmetry spontaneously. The action then can be expanded in the large-$v$ limit, to the order which is relevant to our computation, it is given by, 
\bea
\mathcal{S}_{\rm CDBI} &=& \int d^4 x \left(  -{1 \over 2} (\partial \phi)^2 
- { 1 \over  v^4} {1 \over 2!} \left({(\partial \phi)^2 \over 2} \right)^2 + 
{ 4 \over  v^5} {1 \over 2!} \left({(\partial \phi)^2 \over 2} \right)^2 \phi - { 20 \over v^6}
{1 \over 2!} \left({(\partial \phi)^2 \over 2} \right)^2{\phi^2 \over 2} \right) \cr
&+& \ldots \, .
\eea
Similarly in the case of $D=6$, we have, 
\bea \label{eq:6DCDBI}
\mathcal{S}_{\rm CDBI} &=& \int d^6 x \left(  -{1 \over 2} ( \partial \phi)^2 
- { 1 \over  v^4} {1 \over 2!} \left({(\partial \phi)^2 \over 2} \right)^2 
+ 
{ 6 \over  v^5} {1 \over 2!} \left({(\partial \phi)^2 \over 2} \right)^2 \phi
- 
{ 48 \over v^6}
{1 \over 2!} \left({(\partial \phi)^2 \over 2} \right)^2{\phi^2 \over 2} \right)
\cr
 &+& \ldots \, .
\eea
To obtain the action of $6D$ conformal DBI in the above equation eq.~(\ref{eq:6DCDBI}), we have made a field-redefinition to remove a three-point vertex since it vanishes when on-shell. In the following we compute the UV divergences of one-loop four, five and six-point amplitudes built from these vertices. First we note that the four-point amplitude is identical to that of the flat-space DBI, so the results are given in eq.~(\ref{eq:4pt4D}) and eq.~(\ref{eq:4pt6D}) for the theory at $D=4$ and $D=6$, respectively. Therefore we will only consider five and six-point amplitudes shown in Fig.~\ref{fig:six-ptCDBI}.

\subsubsection{Conformal DBI at $D=4$}
\begin{figure}
\begin{center}
\includegraphics[scale=0.6]{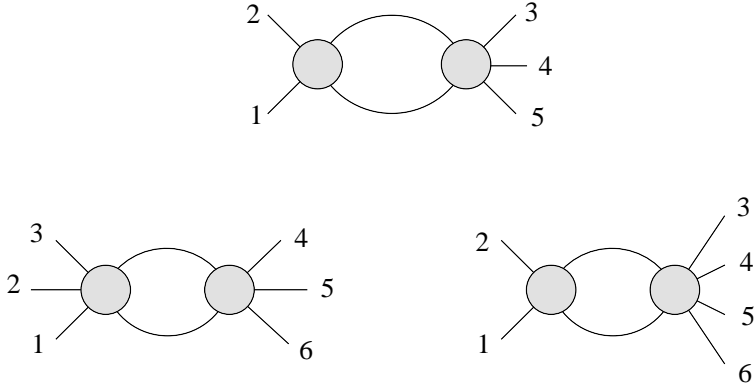}
\caption{ The Feynman diagrams contributing to five and six-point amplitudes in conformal DBI at one loop of order $s^4$ at $D=4$ and $s^5$ at $D=6$, where the four-point vertex is identical to that of DBI, and five- and six-point vertices are 
	$\phi (\partial \phi \cdot \partial \phi )^2$ and $\phi^2 (\partial \phi \cdot \partial \phi )^2$. One should also sum over all other independent permutations.  }
\label{fig:six-ptCDBI}
\end{center}
\end{figure}

The computation is very similar to that of flat space DBI, so we will be brief here, and only present the final results. At $D=4$, the UV divergence of the amplitudes presented in Fig.~\ref{fig:six-ptCDBI} goes as $s^4$. As usual to express the results in a compact form we will use independent polynomial basis.  The basis relevant for the five and six-point UV divergence of $D=4$ conformal DBI we will use is given by, 
\bea
b^{(4)}_{5,1} &=& (s_{12}^2 + \mathcal{P}_5)^2 \, , \quad 
b^{(4)}_{5,2} = s_{12}^4 + \mathcal{P}_5  \cr
b^{(4)}_{6,1} &=& s_{12}^4 + \mathcal{P}_6 \, , \quad 
b^{(4)}_{6,2} = (s_{12}^2 + \mathcal{P}_6)^2 \, , \cr
b^{(4)}_{6,3} &=& s_{12}^2 s_{23}^2 + \mathcal{P}_6 \, , \quad
b^{(4)}_{6,4} =  s_{123}^4 + \mathcal{P}_6 \, , 
\eea
and we denote the six-point ones as a list $B^{(4)}_6 = \{b^{(4)}_{6,1}, b^{(4)}_{6,2}, b^{(4)}_{6,3}, b^{(4)}_{6,4}\}$. Performing the one-loop integral, we then obtain the final results of five- and six-point UV divergences of conformal DBI, 
\bea
A^{(4)}_{5, {\rm UV}} &=&   -{ 43  \pi^2  \over 6480 \epsilon } b^{(4)}_{5,1}
-  { 4 \pi^2 \over 27 \epsilon } b^{(4)}_{5,2}  \,, \\ \nonumber
A^{(4)}_{6, {\rm UV}} &=&  C^{(4)}_{6,1} \cdot B^{(4)}_6 + C^{(4)}_{6,2} \cdot B^{(4)}_6   \, ,
\eea
with the coefficients of six-point case given as
\bea
  \epsilon C^{(4)}_{6,1}  &=& \biggr\{ -{13 \pi^2  \over 270}, {\pi^2 \over 384}, -{49 \pi^2 \over 90}, {163 \pi^2  \over 1620} \biggr\} \, ,
 \\ \nonumber
  \epsilon C^{(4)}_{6,2}  &=& \biggr\{ {2 \pi^2 \over 27}, {25 \pi^2  \over 6912}, -{8 \pi^2 \over 9}, {2 \pi^2 \over 81} \biggr\} \, . 
\eea
We have verified that the results satisfy all the single and double soft theorems of (broken) conformal symmetry. 

\subsubsection{Conformal DBI at $D=6$}
 
Let us now move on the case of conformal DBI at $D=6$. The UV divergence for the amplitudes in Fig. \ref{fig:six-ptCDBI} should go as $s^5$, and the polynomial basis for five-point amplitude we will use are given as: 
\bea
b^{(5)}_{5,1} &=& s_{12}^2 s_{34}^3 + \mathcal{P}_5 \, , \quad 
b^{(5)}_{5,2} = s_{12}^2 s_{23}^3 + \mathcal{P}_5   \, ,
\eea
whereas the polynomial basis for six-point kinematics is the same as that in eq.~(\ref{eq:basiss5}): $B^{(5)}_6 = \{b^{(5)}_{6,1}, b^{(5)}_{6,2}, b^{(5)}_{6,3}, b^{(5)}_{6,4}, b^{(5)}_{6,5}\}$. Note now there is no factorization term. 
In terms of the above polynomial basis, the final  results of UV divergences of five and six-point amplitudes take the following form,  
\bea
A^{(5)}_{5, {\rm UV}} &=&   { 89 \pi^3  \over 280 \epsilon } b^{(5)}_{5,1}
+  { 447 \pi^3 \over 280 \epsilon} b^{(5)}_{5,2}  \,, \\ \nonumber
A^{(5)}_{6, {\rm UV}} &=& C^{(5)}_{6,1} \cdot B^{(5)}_6 + C^{(5)}_{6,2} \cdot B^{(5)}_6  \, ,
\eea
with the coefficients $C^{(5)}_{6,1}$ and $C^{(5)}_{6,2}$ for the six-point case given by
\bea
 \epsilon C^{(5)}_{6,1}  &=& \biggr\{ -{267 \pi^3 \over 280}, {89 \pi^3 \over 84}, -{1781 \pi^3\over 12600}, {309 \pi^3\over 280}, 
 -{13 \pi^3 \over 80}, { \pi^3\over 4200} \biggr\} \, , \\ \nonumber
 \epsilon C^{(5)}_{6,2}  &=& \biggr\{ -{89\pi^3 \over 70}, {96 \pi^3\over 35}, -{89 \pi^3 \over 1050}, {89 \pi^3 \over 70}, 
  -{ 4\pi^3 \over 5}, -{89 \pi^3 \over 525} \biggr\} \, .
\eea
We have again explicitly verified that the soft theorems that can be derived from current algebra are respected by the above one-loop UV divergence.  

\subsection{A-V and K-S action} 
\begin{figure}
\begin{center}
\includegraphics[scale=0.5]{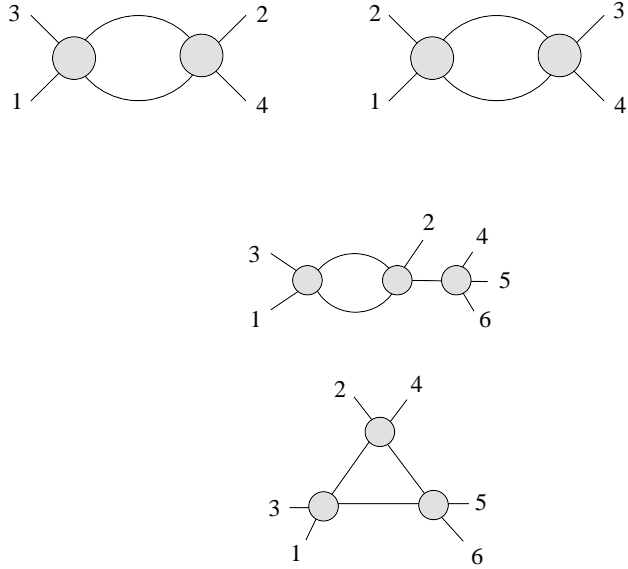}
\caption{ The Feynman diagrams contributing to four- and six-point fermionic amplitudes in K-S action at one loop. One should also sum over all other independent permutations. }
\label{fig:six-ptFermion}
\end{center}
\end{figure}		

In the previous sections we studied the UV divergence of scalar theories, here we consider the A-V model of Glodstino's of spontaneously breaking of supersymmetry. The A-V action takes the form, 
\bea
S_{\rm AV}= - { 1 \over 2 g^2} \int  d^4 x \, {\rm det}  \left( 1 + i g^2 \psi \sigma^{\mu} \overset{\leftrightarrow}{\partial}_{\mu} \bar{\psi} \right) \, .
\eea
One may expand the determinant to obtain all the higher-dimensional operators. It is known that the expansion terminates at the order of $(\psi \bar{\psi})^3$, namely higher orders with eight fermions are actually absent~\cite{Kuzenko:2005wh}. The soft theorems of tree-level scattering amplitudes in Akulov-Volkov theory were established in~\cite{deWit:1975xci, Chen:2014xoa}, and reproduced via current algebra in appendix~\ref{Qbreak}, where it was shown that the amplitudes for the A-V model have the Alder's zero as pions, furthermore the amplitudes also satisfy the double-soft theorems which reflect the underlying supersymmetry algebra.\footnote{Recent study on the soft theorems of amplitudes in A-V theory can be found in~\cite{Kallosh:2016qvo, Kallosh:2016lwj}.} Explicitly, the Adler's zero of amplitudes in A-V theory is, 
\bea
A_{n}( \psi_1, \bar{\psi}_2, \ldots,  \psi_{n-1}, \bar{\psi}_n ){\big{|}}_{{\lambda}_n \rightarrow \tau {\lambda}_n} \sim \mathcal{O}(\tau) \,. 
\eea
Here we also use the standard spinor helicity formalism for massless momenta, 
\bea
p^{\mu} \sigma^{\alpha \dot{\alpha} }_{\mu} = \lambda^{\alpha} \tilde{\lambda}^{\dot{\alpha}}\,,
\quad 
\langle i\,j \rangle = \epsilon_{\alpha\,\beta} \lambda_i^{\alpha} \lambda_j^{\beta}  \, ,
\quad
[ i\,j ] = \epsilon_{\dot{\alpha} \,\dot{\beta}} \tilde{\lambda}_i^{\dot{\alpha}} 
\tilde{\lambda}_j^{\dot{\beta}}
\eea 
and the soft limit of particle $\bar{\psi}_n$ can be realized by setting ${\lambda}_n \rightarrow \tau {\lambda}_n$.\footnote{Here we made a choice such that the soft limits do not rescale the wave functions of fermions, although it is not necessary.} While the double-soft theorem is given by
\bea
A_{n}( \psi_1, \bar{\psi}_2, \ldots,  \psi_{n+1}, \bar{\psi}_{n+2} ){\big{|}}_{
\tilde{\lambda}_{n+1} \rightarrow \tau  \tilde{\lambda}_{n+1}, {\lambda}_{n+2} \rightarrow \tau {\lambda}_{n+2}}
&=& \sum_{i=1}^{n} S_{{\rm F}, i} A_{n}( \psi_1, \bar{\psi}_2, \ldots, \bar{\psi}_{n}) 
\cr 
&+& \mathcal{O}(\tau) \, ,
\eea
with the soft factor given by 
\bea
S_{{\rm F}, i} = {k_i \cdot (k_{n+1}- k_{n+2}) \over 2 k_i \cdot (k_{n+1}+ k_{n+2})} \langle n{+}1| k_i | n{+}2 ] \, ,
\eea
here we defined $\langle i | k_j | l ] :=\langle i\,j\rangle[j\, l]$. The derivation of the above soft theorem using the current algebra of breaking supersymmetry can be found in appendix~\ref{Qbreak}. In the following we test the single and double soft theorems by computing one-loop UV divergence in the A-V theory. 

For the computation of six-point amplitudes at one loop is actually more convenient to use the equivalent Komargodski-Seiberg (K-S) action~\cite{Komargodski:2009rz}, which is related to the A-V action by a non-linear change of variables~\cite{Kuzenko:2010ef}. Furthermore it was shown that~\cite{Kuzenko:2010ef} the K-S action instead does not contain the six-point vertex, but the eight-fermion interaction is now present which however does not contribute to the six-point amplitudes at one loop anyway. So due to the absence of the six-fermion term, it is clear that K-S action simplifies the computation of one-loop six-point amplitude. Explicitly, the K-S action is given by, 
\bea
S_{\rm KS} = \int d^4 x \left( i \partial_{\mu} \bar{\psi} \sigma^{\mu}  \psi
+ {g^2 \over 4}  \bar{\psi}^2 \partial_{\mu} \left( \psi \partial^{\mu} \psi  \right)  + \ldots \right) \, ,
\eea
where we have omitted the eight-fermion term since as we mentioned that it is irrelevant to the computation we are interested in. 
The scattering amplitudes in the theory can be obtained by gluing these four- and eight-point vertices. For the computation in paper, we will only need the four-point vertex, which is given by, 
\bea
V_{\rm F}(\psi_1, \psi_2, \bar{\psi}_3, \bar{\psi}_4) = g^2  \psi_1 \cdot \psi_2 \bar{\psi}_3 \cdot \bar{\psi}_4    (k_1 + k_2 )^2 \, ,
\eea 
when on-shell it reduces to $\langle 12\rangle [34]s_{12}$ that is the four-point amplitude in A-V theory.  

Let us now consider the one-loop fermionic amplitudes in the K-S action. Begin with the four-point case, there are two kinds of Feynman diagrams which can contribute, as shown in Fig.~\ref{fig:six-ptFermion}. To be concrete we consider the amplitude $A_4 (\psi_1, \bar{\psi}_2, \psi_3, \bar{\psi}_4)$, and the one-loop bubble integrands of two diagrams for the four-point amplitude take the form, 
\bea
I^{(1)}_4 = {1 \over 4} { \mathcal{N}_1  \over \ell^2 (\ell + k_1 + k_3) } \,, \quad
I^{(2)}_4 = {1 \over 4} { \mathcal{N}_2  \over \ell^2 (\ell + k_1 + k_2) }
\eea
with numerators are given by
\bea
 \mathcal{N}_1 &=& V_{\rm F}({\psi}_1, {\psi}_3, \bar{\psi}_I, \bar{\psi}_{I'})
 V_{\rm F}({\psi}_I, {\psi}_{I'}, \bar{\psi}_2, \bar{\psi}_4)  \,, 
 \cr
 \mathcal{N}_2 &=& V_{\rm F}({\psi}_1, \bar{\psi}_2, \psi_I, \bar{\psi}_{I'})
 V_{\rm F}(\psi_{I'}, \bar{\psi}_{I}, \psi_3, \bar{\psi}_4)  \,.
\eea
Here $\bar{\psi}_I$ and $\bar{\psi}_{I'}$ are off-shell internal lines. Explicitly, 
\bea
\mathcal{N}_1 &=& 2\langle 1\,3 \rangle [2\,4] s_{13}^2 \ell \cdot (\ell + k_1 + k_3)  \, , \cr
\mathcal{N}_2 &=& \langle 1| \ell_2 |4] \langle 3| \ell|2 ]  (\ell + k_2)^2   (\ell - k_3)^2   \, ,
\eea
where $\ell_2= -(\ell + k_1 + k_2)$ in the expression of $\mathcal{N}_2$. 
Perform the integral, we find the UV divergent part of the first diagram is given by, 
\bea \label{eq:4ptFermion1}
A^{(1)}_{4, {\rm UV}} = {\pi^2 \over 2 \epsilon}   \langle 1\,3 \rangle [2\,4] s_{13}^3 \, ,
\eea
where we have summed over the relevant permutations. While for the second diagram we have,
\bea \label{eq:4ptFermion2}
A^{(2)}_{4, {\rm UV}} =  { \pi^2  \over 120 \epsilon } 
\langle 1\,3 \rangle [2\,4] \left(12s_{13}^3 - 11 (s_{12}^3 + s_{23}^3 +s_{13}^3)   \right) \, .
\eea
Thus the full UV divergence of four-point fermionic amplitude in A-V and K-S theory is given by the sum of the above two contributions. 

Let us now move on to the computation of six-point amplitude
$A_6 (\psi_1, \bar{\psi}_2, \psi_3, \bar{\psi}_4,  \psi_5,  \bar{\psi}_6)$. The Feynman diagrams that contribute to this amplitude are shown in Fig.~\ref{fig:six-ptFermion}, which are obtained by gluing three four-point vertices $V_{\rm F}$. There are six types of Feynman diagrams according to the different assignments of the helcities of fermions. Here we only write explicit integrands for a couple of  Feynman diagrams as examples. For instance, the integrand for the bubble diagram takes the form, 
\bea
I^{(1)}_{\rm B} = {1 \over 2} { \mathcal{N}^{(1)}_{\rm B} \over \ell^2 (\ell + k_1 + k_3) s_{456}  } \,, 
\eea
where the numerator 
\bea
\mathcal{N}^{(1)}_{\rm B} = 
2s_{13}^2 s_{46} \langle 1\,3 \rangle [4\,6] \ell_1 \cdot (\ell_1 + k_1 + k_3) \langle  5|4+6 |2] \,. 
\eea
While the integrand for the triangle diagram is given by, 
\bea
I^{(1)}_{\rm T} = {1 \over 3!} { \mathcal{N}^{(1)}_{\rm T} \over \ell^2 (\ell + k_1 + k_3)^2  (\ell - k_2 - k_4)^2 } \,, 
\eea
where the numerator 
\bea
\mathcal{N}^{(1)}_{\rm T} = 
s_{13} s_{24} \langle 1\,3 \rangle  [2\,4] \langle 5| \ell_3 \ell_1 \ell_2| 6]  (\ell_2 - k_6)^2   \,,
\eea
where $\ell_i$'s are defined as $\ell_1= \ell$, $\ell_2 = \ell- k_2 - k_4$ and $\ell_3= \ell + k_1 + k_3$. The one-loop integration is again straightforward to preform, however unlike the scalar amplitudes of DBI or the four-point fermionic amplitude, the result of the UV divergent part of the six-point fermionic amplitude is rather lengthy.
So we will not present the explicit result here but an auxiliary {\it mathematica} notebook containing the full expression is attached to the arXiv submission. Most importantly we have verified that the one-loop UV divergence of the six-point amplitude in the theory we obtained satisfies the expected vanishing single limit, and a double soft theorem that is consistent with the four-point result in eq.~(\ref{eq:4ptFermion1}) and eq.~(\ref{eq:4ptFermion2}).

\section{Super and bosonic string amplitudes}
\label{sec:strings}
The massless sector of open string theory, when dimensionally reduced to $p{+}1$-dimensions, corresponds to the low energy degrees of freedom of a stack of $D$-$p$ branes. For example, the six scalars of four-dimensional  $\mathcal{N}=4$ super-Yang mills are the Goldstone modes for the broken translation symmetry in the transverse directions of $D$-$3$ branes. Thus the open bosonic and superstring amplitudes encode the information of two distinct D-brane effective actions, which should satisfy all soft theorems derivable from broken translational symmetry. In this section we will consider only the pure scalar part of the effective action.

Isolating the interactions of the centre of mass degrees of freedom for the branes correspond to separating $U(N)\rightarrow U(1)\times SU(N)$, and keeping only the U(1) part. This is done in practice by summing over all orderings of the open-string amplitude. In the end, one obtains an on-shell effective action with the schematic form:
\eq
\mathcal{L}= - {1 \over 2}\phi \Box \phi+\sum_{m, 3<n}c_{n,m}\alpha'^{\frac{n}{2}-2+m}\p^{2m} \phi^n \,.
\eqe
Here, we will translate $\p^{2m} \phi^n$ into momentum space where they become combinations of permutation invariant polynomials of $s_{i,j}$. The coefficients $c_{n,m}$ are generally given in terms of multiple zeta values (MZVs), defined as:
\eq
\zeta_{n_1,n_2,\cdots,n_r}\equiv\sum_{0<k_1<k_2<\cdots<k_r}^{\infty} \frac{1}{k_1^{n_1}k_2^{n_2}\cdots k_r^{n_r}}\,.
\eqe 
MZVs can be conjecturally categorised according to their transcendental weight $n_1 + n_2 + . . . + n_r$, and for maximally supersymmetric string theories, it is known that the transcendental weight for each coefficient matches the order of $\alpha'$, which is coined as \textit{uniform transcendentality} property. For non-maximal theories, while the leading transcendental pieces match with the maximal case~\cite{Huang:2016tag}, subleading pieces may also be present.

Note that the lowest mass-dimension piece of the $n$-point amplitude (whose local part takes the form $s_{i,j}^{\frac{n}{2}}$) must be identical with DBI. This is because permutation invariance forbids two derivative four-point vertex, while single soft-theorems are sufficient to compeletly determine the lowest dimension amplitudes from the four-derivative quartic vertex, i.e that of DBI ~\cite{Cheung:2015ota, Arkani-Hamed:2016rak}. In fact, this piece is leading transcendental, and hence it should be universal. In this section we will consider higher order corrections in $\alpha'$ for the four- and six-point open string amplitudes. They can be derived by simply taking the gluon amplitudes in string theory, identifying $\epsilon_i\cdot k_j=0$, $\epsilon_i\cdot \epsilon_j=1$ (namely dimension reduction), and summing over all permutations.

As we will show, the scalar modes that are associated with the center of mass of the D-branes will exhibit soft behaviours associated with the spontaneous translation symmetry breaking. 

\subsection{Type $I$ superstring }
The massless amplitudes of type-$I$ superstring can be naturally represented as~\cite{Mafra:2011nv}:
\bea \label{eq:generalopen}
{\cal A}^S(1,\rho_1, \ldots, \rho_{n-2}, n-1,n;\alpha') = 
\sum_{\sigma \in S_{n-3}} F_\rho{}^{\sigma}(\alpha')\nonumber\\
\times A_{\rm YM} (1, 2_{\sigma }, \ldots, (n-2)_{\sigma }, n-1, n) \ ,
\eea
where ${\cal A}^S$ and $A_{\rm YM}$ indicate color-ordered Yang-Mills amplitudes of the superstring and super Yang-Mills field theory, respectively. Moreover, $\rho,\sigma$ labels all  $(n-3)!$ distinct permutations with legs $(2,3,\cdots,n-2)$. The function $F_{\rho}{}^{\sigma}(\alpha')$ are disk integrals with insertion points $(z_1,z_{n-1},z_n)$ fixed to $(0,1,\infty)$ respectively, and
\bea
F_\rho{}^{\sigma}(\alpha') \equiv 
\! \! \! \! \! \! \! \! \! \! \! \! \! \! \! \! \! \! \! \! \! \! \! \! \! \! \int \limits_{0\leq  z_{\rho_1} \leq z_{\rho_2}\leq \ldots \leq z_{\rho_{(n-2)}} \leq 1 }\! \! \! \! \! \! \! \! \! \! \! \! \! \! \! \! \! \! \! \! \! \! \! \! \! \!
  d^2 z_2 \ldots d^2 z_{n-2} \, \prod^n_{i<l} |z_{il}|^{s_{il}} \sigma \Big\{ 
 \prod^{n-2}_{k=2} \sum^{k-1}_{m=1} {s_{mk}  \over z_{km}} \Big\} \, , 
 \label{oli0}
\eea
with $z_{ij} \equiv z_i - z_j$. When viewed as an $(n-3)! \! \times \!(n-3)!$ matrix, the row- and column indices $\rho$ and $\sigma$ of $F_\rho{}^{\sigma}$ label different integration domains and integrands, respectively, where $\sigma$ acts on the subscripts within the curly bracket in eq.~(\ref{oli0}). Note that the field-theory limit is recovered as $F_\rho{}^{\sigma}(\alpha') = \delta_\rho{}^{\sigma} +{\cal O}(\alpha'^2)$. The full $\alpha'$ expansion is conveniently organised as~\cite{Schlotterer:2012ny}
\begin{align}
F(\alpha') &= 1 + \zeta_2 P_2+\zeta_3 M_3+\zeta_2^2 P_4 + \zeta_5 M_5+\zeta_2\zeta_3 P_2M_3 + \zeta_2^3 P_6 + {1 \over 2} \zeta_3^2 M_3 M_3 + \zeta_7 M_7 + \ldots \ ,
\label{oli10}
\end{align}
where the entries of the $(n-3)!\times (n-3)!$ matrices $P_w, M_w$ are degree $w$ polynomials in $\alpha' s_{ij}$ with rational coefficients. The precise forms of these matrices can be found in~\cite{OliverWeb}. 

To obtain the scalar amplitudes, we simply take $A_{\rm YM}$ in eq.~(\ref{eq:generalopen}) and set $\epsilon_i\cdot k_j=0$, $\epsilon_i\cdot \epsilon_j=1$. Summing over all permutations of the external momenta, and taking the $\alpha'$ expansion one finds 
 \eq
 \mathcal{L}^{Type-I}= - {1 \over 2}\phi \Box \phi + \alpha'^2 \pi^2 [\p^4 \phi^4]_s
 + \alpha'^4 \pi^4 [\p^8 \phi^4]_s + \alpha'^3 \pi^3 [\p^6 \phi^6]_s + 
\alpha'^5 \pi^5 [\p^{10} \phi^6]_s + \ldots \, ,
 \eqe
 where we note at four point $[\p^6 \phi^4]_s$ is absent due to supersymmetry. The explicit polynomials represented in $[\p^n \phi^m]_s$ are given as, 
 \bea
  [\p^4 \phi^4]_s = {1 \over 2}(s^2+ t^2 + u^2) \, , \quad
  [\p^8 \phi^4]_s = {1 \over 24}(s^4+ t^4 + u^4)  \, .
 \eea
 At six points, $[\p^6 \phi^6]_s$ is the term corresponding to that of DBI, while $[\p^{8} \phi^6]_s$ has a vanishing coefficient in super string theory. At the order of ten derivatives we have, 
 \bea
[\p^{10} \phi^6]_s &=&- {1 \over 384}b_1^{(5)} - {1 \over 864}b_2^{(5)} - {1 \over 51840}b_3^{(5)} 
-  {1 \over 192}b_4^{(5)}
\cr
&& -  {13 \over 1152}b_5^{(5)} + { 1 \over 1080} b_6^{(5)} \, . 
\eea
where the polynomial basis are given by, 
\bea \label{eq:basis5}
b_1^{(5)} &=& s_{123}^2 s_{234}^3 + \mathcal{P}_6  \, , \quad 
b_2^{(5)} = s_{12}^2 s_{23}^3+ \mathcal{P}_6 \, , \cr
b_3^{(5)} &=& s_{123}^5 + \mathcal{P}_6 \, , \quad 
b_4^{(5)} = s_{34}^2 s_{123}^3 + \mathcal{P}_6 \, ,  \cr
b_5^{(5)} &=&  s_{34}^3 s_{123}^2 + \mathcal{P}_6 \, , \quad 
b_6^{(5)} =  s_{12}^5 + \mathcal{P}_6 \, .
\eea 
Written the effective action in this explicit basis, it is straightforward to verify that the scattering amplitudes (up to six points at order $s^5$) satisfy the leading and sub-leading single-soft translation soft theorems, as well as double soft theorems up to order $\tau^2$. The proposed $\mathcal{O}(\tau^3)$ soft theorem in~\cite{Cachazo:2015ksa} is found not to hold as expected since it cannot be derived from current algebra. As the case of the UV divergence at $D=4$, this is due to the presence of the four-point amplitude of order $s^4$. 

\subsection{Bosonic string }
The derivation of bosonic string result is more involved, and we consider the standard representation of the bosonic string integrand. Taking $\epsilon_i\cdot k_j=0$, $\epsilon_i\cdot \epsilon_j=1$ for the bosonic string one obtains the integrand of the form:
\bea
I_\rho(\alpha') \equiv 
\int \limits_{0\leq  z_{\rho_1} \leq z_{\rho_2}\leq \ldots \leq z_{\rho_{(n-2)}} \leq 1 }  d^2 z_2 \ldots d^2 z_{n-2} \, \prod^n_{i<l} |z_{il}|^{s_{il}}  \Big\{ \frac{1}{z^2_{12}z^{2}_{34}\cdots z^2_{n{-1}n}}
  +\mathcal{P}_n\Big\} \, , 
\eea
The above ``multi trace" integrals can be systematically reduced to ``single traced" ones via integration by parts (IBP) identities. The relevant identities are listed in appendix~\ref{IBP}. Thus the resulting ``scalar" piece of the amplitude is given in the form:
\eq
\sum_{\rho\in S_5}c_{\rho}(s_{i,j})Z(1\rho_2\rho_3\rho_4\rho_5\rho_6)\,,
\eqe
where one sums over all $5!$ permutations of $(2,3,4,5,6)$ labeled by $\rho$, and 
\eq
Z(123456)\equiv \int 
 \left(\prod_{i=2}^5 d^2 z_i\right) \frac{\prod^6_{i<l} |z_{il}|^{s_{il}}}{z_{12}z_{23}z_{34}z_{45}z_{56}z_{61}}\,. 
\eqe
The single trace disk integrand satisfy KK- and BCJ-relations, and thus one can further reduce the representation to that involving only six distinct single trace integrand which can now be cast in terms of the $F_\rho{}^{\sigma}(\alpha')$ of superstring. More precisely, we have~\cite{Broedel:2013tta}
\eqa
Z(6\sigma_2\sigma_3\sigma_451)=&-&\frac{F^{\sigma_2\sigma_3\sigma_4}}{s_{16}}\left(\frac{1}{s_{\sigma_3\sigma_4}s_{\sigma_3\sigma_45}}+\frac{1}{s_{\sigma_45}s_{\sigma_3\sigma_45}}+\frac{1}{s_{\sigma_2\sigma_3}s_{\sigma_45}}+\frac{1}{s_{\sigma_4\sigma_3}s_{156}}+\frac{1}{s_{\sigma_2\sigma_3}s_{156}}\right)\nonumber\\
&+&\frac{F^{\sigma_2\sigma_4\sigma_3}}{s_{16}s_{\sigma_3\sigma_4}}\left(\frac{1}{s_{\sigma_3\sigma_45}}+\frac{1}{s_{156}}\right)+\frac{F^{\sigma_3\sigma_2\sigma_4}}{s_{16}s_{\sigma_2\sigma_3}}\left(\frac{1}{s_{\sigma_45}}+\frac{1}{s_{156}}\right)\nonumber\\
&+&\frac{F^{\sigma_3\sigma_4\sigma_2}}{s_{16}s_{\sigma_3\sigma_4}s_{156}}+\frac{F^{\sigma_4\sigma_2\sigma_3}}{s_{16}s_{\sigma_2\sigma_3}s_{156}}-\frac{F^{\sigma_4\sigma_3\sigma_2}}{s_{16}s_{156}}\left(\frac{1}{s_{\sigma_2\sigma_3}}+\frac{1}{s_{\sigma_4\sigma_3}}\right)
\eqae 
where $\sigma_2\sigma_3\sigma_4$ corresponds to the different permutations of (234), and the functions $F^{\sigma_2\sigma_3\sigma_4}$ are defined with canonical ordering $0\leq z_2\leq z_3\leq z_4\leq 1$. 

Plugging in the explicit $\alpha'$ expansion denoted in eq.~(\ref{oli10}), we find the following effective action for bosonic open string:
\eqa
 \mathcal{L}^{Bosonic}&=& - {1 \over 2}\phi \Box \phi + \pi^2 \alpha'^2 [\p^4 \phi^4]_b + \pi^2 \alpha'^3 [\p^6 \phi^4]_b + \pi^4 \alpha'^4 [\p^8 \phi^4]_b  +  \pi^3 \alpha'^3 [\p^6 \phi^6]_b \nonumber\\
 &+& \pi^3 \alpha'^4 [\p^8 \phi^6]_b +
 \alpha'^5 [\p^{10} \phi^6]_b
+ \ldots \eqae
we note the appearance of non-maximal transcendental terms. Here the four-point vertices take following explicit expressions, 
\bea
[\p^4 \phi^4]_b &=&  {1 \over 2} (s^2 +t^2 +u^2) \, , \quad
[\p^6 \phi^4]_b =(s^3 + t^3 + u^3) \, , \cr
[\p^8 \phi^4]_b &=&  {1 \over 24} (s^4 +t^4 +u^4) \, . 
\eea
Now the eight-derive term is given by
\bea
[\p^8 \phi^6]_b = {1 \over 288} b_1^{(4)} + {1 \over 12} b_2^{(4)} - {1 \over 108} b_3^{(4)} - {3 \over 32} b_4^{(4)} \, ,
\eea 
with following polynomial basis,
 \bea
b_1^{(4)} &=& s_{12}^4 + \mathcal{P}_6 \, , \quad 
b_2^{(4)} = s_{12}^2 s_{23}^2 + \mathcal{P}_6 \cr
b_3^{(4)} &=& s_{123}^4 + \mathcal{P}_6 \, , \quad 
b_4^{(4)} = s_{123}^2 s_{34}^2 + \mathcal{P}_6  \, .
\eea
The ten-derivative term takes the following form,   
\bea
[\p^{10} \phi^6]_b
&=& [\p^{10} \phi^6]_{s}+ \pi^4 \left( - {1 \over 8}b_1^{(5)} +  {13 \over 36}b_2^{(5)} - {1\over 108}b_3^{(5)} 
+ {1 \over 12}b_4^{(5)} \right.
\cr
&& \left. - {11 \over 48}b_5^{(5)} +{ 1 \over 72} b_6^{(5)} \right)
\eea
where the polynomial basis $b_i^{(5)}$ are defined in eq.~(\ref{eq:basis5}), and the term $[\p^{10} \phi^6]_{s}$ is identical to that in the superstring. We note ten-derivative term $\p^{10} \phi^6$ in bosonic string contains lower-transcendental terms. Again we have verified the scattering amplitudes from this effective action satisfy the single and double-soft theorem up to $\mathcal{O}(\tau^2)$. Interestingly, comparing to DBI, the $\alpha'$-expansion of the bosonic string not only generates a four-point vertex with eight derivatives (it is the same as the superstring, and that would violate the double-soft theorems of order $\mathcal{O}(\tau^3)$), but also a six-derivative one. A simple power counting shows that a six-derivative four-point vertex could potentially change the double-soft theorems of tree-level DBI even at the order $\mathcal{O}(\tau^2)$. However as we argued in the previous section~\ref{sec:derivation}, the double-soft theorem at this order is derivable from symmetry principle and should be protected from UV divergence as well as $\alpha'$ corrections, and our explicit computation shows that is indeed the case. Thus the double-soft theorems up to $\mathcal{O}(\tau^2)$ is solely due to enhanced broken symmetries and protected from any possible modification due to new higher dimension operators. 

\section{Conclusion and outlook} \label{sec:conclusion}
In this paper, we study sub-leading soft theorems that arise from the enhanced broken symmetries. This occurs for space-time symmetries, where some generators of the broken symmetries are derivatively related. This then implies that the multiple broken symmetries lead to the same Goldstone mode, and in the soft momenta expansion, the presence of universal behaviour at sub-leading order and beyond. We have applied the analysis to spontaneously broken conformal and translational symmetry, deriving double-soft theorems at leading and sub-leading level. This allows us to identify the $\mathcal{O}(\tau)$ and $\mathcal{O}(\tau^2)$ double-soft theorems given in~\cite{Cachazo:2015ksa}(for DBI), can be attributed to symmetry arguments alone, while the order-$\mathcal{O}(\tau^3)$ soft-theorem requires particular quartic interactions.

Naively one would expect that the use of currents $(j_{K}, j_{K})$ of conformal boost should lead to a sub-sub-leading double soft theorem of dilatons. However, that cannot be the case.\footnote{A universal factorized sub-sub-leading double soft theorem of dilatons may not exist has also suggested in \cite{DiVecchia:2017uqn}. }  One can also see this by studying the amplitude of six dilatons at the order of $s^3$. This amplitude is completely fixed by the dilaton single soft theorem, and the result can be expressed in terms of lower-point amplitudes, namely four-point order-$s^2$ and five-point order-$s^3$ amplitudes. If a sub-sub-leading double soft theorem exists, it would mean that in the double-soft limit, at this order the six-point amplitude should be proportional to the four-point order-$s^2$ amplitude, but from explicit computation we found that is not the case. Thus there cannot be such a universal double-soft theorem. 

These soft theorems are expected to be exact, and we give explicit tests against UV divergences of effective field theories whose tree-level (classical action) amplitudes have the right soft behaviours. In particular we have computed and tested
\begin{itemize}
  \item The one-loop UV divergences for DBI action at four- and six-points in $D=4,6, 8$: for single-soft theorems up to $\mathcal{O}(\tau)$, and double-soft theorems up to $\mathcal{O}(\tau^2)$
  \item The one-loop UV divergences for conformal DBI action at four-, five- and six-points in $D=4,6$, for single-soft theorems up to $\mathcal{O}(\tau)$, and double-soft theorems up to $\mathcal{O}(\tau^2)$.
  \item The one-loop UV divergence for A-V model at four- and six-points in $D=4$ for double-soft theorems at $\mathcal{O}(\tau^0)$,
\end{itemize}   
where all soft-theorems are satisfied. This lends support to the statement that soft-theorems derived from current algebras, which are equivalent to non-linear symmetries, are exact even for the regulated theory. We also consider the S-matrix of open string effective field theory, for which the translation symmetry breaking induced double-soft theorems are also shown to hold up to $\mathcal{O}(\tau^2)$. More over, operators that are power counting capable of modifying the $\mathcal{O}(\tau^2)$ soft-theorem are present for the bosonic string, thus the fact that it is still preserved reflects the non-triviality of such symmetry based soft-theorems. It is quite remarkable how weakly coupled string amplitudes know about the presence of D-branes in such non-trivial fashion.

Note that the soft-theorems for the D-brane effective field theory changes depending on the isometries of the back ground, as demonstrated for flat space and AdS. It would be interesting to consider other non-trivial backgrounds, to derive the associated soft-theorems and in turn constrain its effective action. These can be useful in consider the effective action for more complicated Coloumb branches. It has been shown that soft theorems in itself for NLSM and DBI are sufficient to enforce unitarity~\cite{Arkani-Hamed:2016rak, Rodina:2016mbk, Rodina:2016jyz}, and it would be nice to show that this continues to be true for conformal DBI or any other new backgrounds. It was understood recently that the soft theorems of NLSM and DBI are inherent from Weinberg's soft theorems of YM and gravity due to unifying relations among these theories at tree level~\cite{Cheung:2017ems}, it would be of interest to study the implications of our results on the unifying relations.

\section{Acknowledgements}
We thank Massimo Bianchi, and Renata Kallosh for helpful discussion. 
Y-t Huang and Zhizhong Li are supported by MOST under the grant No. 103-2112-M-002-025-MY3 and the support from National Center for Theoretical Science (NCTS), Taiwan. ALG is supported by the S\~ao Paulo Research Foundation (FAPESP) under grants 2016/01343-7 and 2017/03303-1, and  by the CUniverse research promotion project by Chulalongkorn University (grant reference CUAASC). The work of CW is supported in part by a DOE Early Career Award under Grant No. DE-SC0010255. 

\appendix 
\section{Broken Supersymmetry}\label{Qbreak}
Using current algebra, we may also get double soft theorems for Goldstino of broken supersymmetry.  Consider the correlator of $2n+2$ currents, by the Ward identity,\footnote{Note that once a fermionic generator, say $Q$, pass through a fermionic current, there would be an extra $-1$.}
\eqa
&&{\rm LSZ} \int dy \, e^{iqy} {\p \over \p y^\nu} \int dx \, e^{ipx} {\p \over \p x^\mu} \langle j_{Q_\alpha}^\mu (x) j_{\bar{Q}_{\dot{\beta}}}^\nu (y) j_{Q_{\gamma_1}}^{\sigma_1} (x_1) \cdots j_{Q_{\gamma_n}}^{\sigma_n} (x_n) j_{\bar{Q}_{{\dot{\gamma}_{n+1}}}}^{\sigma_{n+1}} (x_{n+1}) \cdots j_{\bar{Q}_{{\dot{\gamma}_{2n}}}}^{\sigma_{2n}} (x_{2n}) \rangle \nonumber\\
&=& {\rm LSZ} \int dy \, e^{iqy} {\p \over \p y^\nu} \Bigg[e^{ipy} \langle j_{P_{\alpha \dot{\beta}}}^\nu (y) j_{Q_{\gamma_1}}^{\sigma_1} (x_1) \cdots j_{Q_{\gamma_n}}^{\sigma_n} (x_n) j_{\bar{Q}_{{\dot{\gamma}_{n+1}}}}^{\sigma_{n+1}} (x_{n+1}) \cdots j_{\bar{Q}_{{\dot{\gamma}_{2n}}}}^{\sigma_{2n}} (x_{2n}) \rangle \nonumber\\
&&+\sum_{i=1}^n e^{ipx_{n+i}} (-)^{n+i} \langle j_{\bar{Q}_{\dot{\beta}}}^\nu (y) j_{Q_{\gamma_1}}^{\sigma_1} (x_1) \cdots j_{Q_{\gamma_n}}^{\sigma_n} (x_n) j_{\bar{Q}_{{\dot{\gamma}_{n+1}}}}^{\sigma_{n+1}} (x_{n+1}) \cdots j_{P_{\alpha \dot{\gamma}_{n+i}}}^{\sigma_{n+i}} (x_{n+i}) \cdots j_{\bar{Q}_{{\dot{\gamma}_{2n}}}}^{\sigma_{2n}} (x_{2n}) \rangle \Bigg] \nonumber\\
&=& {\rm LSZ}\, (-iq_\nu) \int dy \, e^{i(p+q)y} \langle j_{P_{\alpha \dot{\beta}}}^\nu (y) j_{Q_{\gamma_1}}^{\sigma_1} (x_1) \cdots j_{Q_{\gamma_n}}^{\sigma_n} (x_n) j_{\bar{Q}_{{\dot{\gamma}_{n+1}}}}^{\sigma_{n+1}} (x_{n+1}) \cdots j_{\bar{Q}_{{\dot{\gamma}_{2n}}}}^{\sigma_{2n}} (x_{2n}) \rangle \nonumber\\
&&+\sum_{i, j=1}^n e^{i(px_{n+i}+qx_j)} (-)^{n+i+j-1}\nonumber\\
&&\times \langle j_{Q_{\gamma_1}}^{\sigma_1} (x_1) \cdots j_{P_{\gamma_j \dot{\beta}}}^{\sigma_j} (x_j) \cdots j_{Q_{\gamma_n}}^{\sigma_n} (x_n) j_{\bar{Q}_{{\dot{\gamma}_{n+1}}}}^{\sigma_{n+1}} (x_{n+1}) \cdots j_{P_{\alpha \dot{\gamma}_{n+i}}}^{\sigma_{n+i}} (x_{n+i}) \cdots j_{\bar{Q}_{{\dot{\gamma}_{2n}}}}^{\sigma_{2n}} (x_{2n}) \rangle \label{super}
\eqae
On the other hand, by exchanging the first two currents and the order of integration in the LHS, we get
\eqa
&&{\rm LSZ} \int dy \, e^{iqy} {\p \over \p y^\nu} \int dx \, e^{ipx} {\p \over \p x^\mu} \langle j_{Q_\alpha}^\mu (x) j_{\bar{Q}_{\dot{\beta}}}^\nu (y) j_{Q_{\gamma_1}}^{\sigma_1} (x_1) \cdots j_{Q_{\gamma_n}}^{\sigma_n} (x_n) j_{\bar{Q}_{{\dot{\gamma}_{n+1}}}}^{\sigma_{n+1}} (x_{n+1}) \cdots j_{\bar{Q}_{{\dot{\gamma}_{2n}}}}^{\sigma_{2n}} (x_{2n}) \rangle \nonumber\\
&=&-{\rm LSZ} \int dx \, e^{ipx} {\p \over \p x^\mu} \int dy \, e^{iqy} {\p \over \p y^\nu} \langle j_{\bar{Q}_{\dot{\beta}}}^\nu (y) j_{Q_\alpha}^\mu (x) j_{Q_{\gamma_1}}^{\sigma_1} (x_1) \cdots j_{Q_{\gamma_n}}^{\sigma_n} (x_n) j_{\bar{Q}_{{\dot{\gamma}_{n+1}}}}^{\sigma_{n+1}} (x_{n+1}) \cdots j_{\bar{Q}_{{\dot{\gamma}_{2n}}}}^{\sigma_{2n}} (x_{2n}) \rangle \nonumber\\
&=& {\rm LSZ} (ip_\mu) \int dx \, e^{i(p+q)x} \langle j_{P_{\alpha \dot{\beta}}}^\mu (x) j_{Q_{\gamma_1}}^{\sigma_1} (x_1) \cdots j_{Q_{\gamma_n}}^{\sigma_n} (x_n) j_{\bar{Q}_{{\dot{\gamma}_{n+1}}}}^{\sigma_{n+1}} (x_{n+1}) \cdots j_{\bar{Q}_{{\dot{\gamma}_{2n}}}}^{\sigma_{2n}} (x_{2n}) \rangle \nonumber\\
&&-\sum_{i, j=1}^n e^{i(px_{n+i}+qx_j)} (-)^{n+i+j-1}\nonumber\\
&&\times \langle j_{Q_{\gamma_1}}^{\sigma_1} (x_1) \cdots j_{P_{\gamma_j \dot{\beta}}}^{\sigma_j} (x_j) \cdots j_{Q_{\gamma_n}}^{\sigma_n} (x_n) j_{\bar{Q}_{{\dot{\gamma}_{n+1}}}}^{\sigma_{n+1}} (x_{n+1}) \cdots j_{P_{\alpha \dot{\gamma}_{n+i}}}^{\sigma_{n+i}} (x_{n+i}) \cdots j_{\bar{Q}_{{\dot{\gamma}_{2n}}}}^{\sigma_{2n}} (x_{2n}) \rangle \label{super2}
\eqae
Summing eq.~(\ref{super}) and eq.~(\ref{super2}) and dividing them by $2$, we have
\eqa
&&{\rm LSZ} \int dy \, e^{iqy} {\p \over \p y^\nu} \int dx \, e^{ipx} {\p \over \p x^\mu} \langle j_{Q_\alpha}^\mu (x) j_{\bar{Q}_{\dot{\beta}}}^\nu (y) j_{Q_{\gamma_1}}^{\sigma_1} (x_1) \cdots j_{Q_{\gamma_n}}^{\sigma_n} (x_n) j_{\bar{Q}_{{\dot{\gamma}_{n+1}}}}^{\sigma_{n+1}} (x_{n+1}) \cdots j_{\bar{Q}_{{\dot{\gamma}_{2n}}}}^{\sigma_{2n}} (x_{2n}) \rangle \nonumber\\
&=& {\rm LSZ}\, {i(p-q)_\mu \over 2} \int dx \, e^{i(p+q)x} \langle j_{P_{\alpha \dot{\beta}}}^\mu (x) j_{Q_{\gamma_1}}^{\sigma_1} (x_1) \cdots j_{Q_{\gamma_n}}^{\sigma_n} (x_n) j_{\bar{Q}_{{\dot{\gamma}_{n+1}}}}^{\sigma_{n+1}} (x_{n+1}) \cdots j_{\bar{Q}_{{\dot{\gamma}_{2n}}}}^{\sigma_{2n}} (x_{2n}) \rangle \,.
\eqae
Then we proceed by performing the LSZ reduction on the both sides, which gives
\eqa
&&\left.M(v(p) \bar{v}(q) v(k_1) \cdots v(k_n) \bar{v}(k_{n+1}) \cdots \bar{v}(k_{2n}))\right|_{p,q\rightarrow0} \\
&=& \sum_{i=1}^{2n} {k_i \cdot (p-q) \over 2 k_i \cdot (p+q)} \langle v(p) | k_i | \bar{v}(q) \rangle M(v(k_1) \cdots v(k_n) \bar{v}(k_{n+1}) \cdots \bar{v}(k_{2n})) + O(p^2, p\cdot q,q^2),\nonumber
\eqae
where we have used that $\langle v(k_i)|j_{P_{\alpha \dot{\beta}}}^\mu(p)|\bar{v}(k_i+p)\rangle= k_{i,\alpha \dot{\beta}}k_i^\mu+\mathcal{O}(p)$.

\section{Broken Supercomformal Symmetry}\label{Sbreak}
For double soft theorem of Goldstino of broken superconformal symmetry, we can instead consider:
\eqa
&&{\rm LSZ} \int dy \, e^{iqy} {\p \over \p y^\nu} \int dx \, e^{ipx} {\p \over \p x^\mu} \langle j_{S_\alpha}^\mu (x) j_{\bar{S}_{\dot{\beta}}}^\nu (y) j_{S_{\gamma_1}}^{\sigma_1} (x_1) \cdots j_{S_{\gamma_n}}^{\sigma_n} (x_n) j_{\bar{S}_{{\dot{\gamma}_{n+1}}}}^{\sigma_{n+1}} (x_{n+1}) \cdots j_{\bar{S}_{{\dot{\gamma}_{2n}}}}^{\sigma_{2n}} (x_{2n}) \rangle \nonumber\\
&=& {\rm LSZ} \int dy \, e^{iqy} {\p \over \p y^\nu} \Bigg[e^{ipy} \langle j_{K_{\alpha \dot{\beta}}}^\nu (y) j_{S_{\gamma_1}}^{\sigma_1} (x_1) \cdots j_{S_{\gamma_n}}^{\sigma_n} (x_n) j_{\bar{S}_{{\dot{\gamma}_{n+1}}}}^{\sigma_{n+1}} (x_{n+1}) \cdots j_{\bar{S}_{{\dot{\gamma}_{2n}}}}^{\sigma_{2n}} (x_{2n}) \rangle \nonumber\\
&&+\sum_{i=1}^n e^{ipx_{n+i}} (-)^{n+i} \langle j_{\bar{S}_{\dot{\beta}}}^\nu (y) j_{S_{\gamma_1}}^{\sigma_1} (x_1) \cdots j_{S_{\gamma_n}}^{\sigma_n} (x_n) j_{\bar{S}_{{\dot{\gamma}_{n+1}}}}^{\sigma_{n+1}} (x_{n+1}) \cdots j_{K_{\alpha \dot{\gamma}_{n+i}}}^{\sigma_{n+i}} (x_{n+i}) \cdots j_{\bar{S}_{{\dot{\gamma}_{2n}}}}^{\sigma_{2n}} (x_{2n}) \rangle \Bigg] \nonumber\\
&=& {\rm LSZ}\, (-iq_\nu) \int dy \, e^{i(p+q)y} \langle j_{K_{\alpha \dot{\beta}}}^\nu (y) j_{S_{\gamma_1}}^{\sigma_1} (x_1) \cdots j_{S_{\gamma_n}}^{\sigma_n} (x_n) j_{\bar{S}_{{\dot{\gamma}_{n+1}}}}^{\sigma_{n+1}} (x_{n+1}) \cdots j_{\bar{S}_{{\dot{\gamma}_{2n}}}}^{\sigma_{2n}} (x_{2n}) \rangle \nonumber\\
&&+\sum_{i, j=1}^n e^{i(px_{n+i}+qx_j)} (-)^{n+i+j-1}\nonumber\\
&&\times \langle j_{S_{\gamma_1}}^{\sigma_1} (x_1) \cdots j_{K_{\gamma_j \dot{\beta}}}^{\sigma_j} (x_j) \cdots j_{S_{\gamma_n}}^{\sigma_n} (x_n) j_{\bar{S}_{{\dot{\gamma}_{n+1}}}}^{\sigma_{n+1}} (x_{n+1}) \cdots j_{K_{\alpha \dot{\gamma}_{n+i}}}^{\sigma_{n+i}} (x_{n+i}) \cdots j_{\bar{S}_{{\dot{\gamma}_{2n}}}}^{\sigma_{2n}} (x_{2n}) \rangle \label{super}
\eqae
Exchange the first two currents and the order of integration in the LHS, and then average them like the previous discussion, we have
\eqa
&&{\rm LSZ} \int dy \, e^{iqy} {\p \over \p y^\nu} \int dx \, e^{ipx} {\p \over \p x^\mu} \langle j_{S_\alpha}^\mu (x) j_{\bar{S}_{\dot{\beta}}}^\nu (y) j_{S_{\gamma_1}}^{\sigma_1} (x_1) \cdots j_{S_{\gamma_n}}^{\sigma_n} (x_n) j_{\bar{S}_{{\dot{\gamma}_{n+1}}}}^{\sigma_{n+1}} (x_{n+1}) \cdots j_{\bar{S}_{{\dot{\gamma}_{2n}}}}^{\sigma_{2n}} (x_{2n}) \rangle \nonumber\\
&=& {\rm LSZ} \, {i(p-q)_\mu \over 2} \int dx \, e^{i(p+q)x} \langle j_{K_{\alpha \dot{\beta}}}^\mu (x) j_{S_{\gamma_1}}^{\sigma_1} (x_1) \cdots j_{S_{\gamma_n}}^{\sigma_n} (x_n) j_{\bar{S}_{{\dot{\gamma}_{n+1}}}}^{\sigma_{n+1}} (x_{n+1}) \cdots j_{\bar{S}_{{\dot{\gamma}_{2n}}}}^{\sigma_{2n}} (x_{2n}) \rangle 
\eqae
Finally apply the LSZ reduction on the both sides, we obtain
\eqa
&&\left.M(v(p) \bar{v}(q) v(k_1) \cdots v(k_n) \bar{v}(k_{n+1}) \cdots \bar{v}(k_{2n})) \right|_{p,q\rightarrow0} \\
&=& \sum_{i=1}^{2n} {k_i \cdot (p-q) \over 2 k_i \cdot (p+q)} \langle v(p) | K_i | \bar{v}(q) \rangle M(v(k_1) \cdots v(k_n) \bar{v}(k_{n+1}) \cdots \bar{v}(k_{2n})) + O(p^2, p\cdot q,q^2),\nonumber
\eqae
where we have used that $\langle v(k_i)|j_{K_{\alpha \dot{\beta}}}^\mu(p)|\bar{v}(k_i+p)\rangle= K_{\alpha \dot{\beta}}k_i^\mu+\mathcal{O}(p)$, and $K_i$ stands for the conformal boost operator acting on the $i$-th particle.
\section{IBP relations for string amplitudes \label{IBP}}
Here we list the relevant IBP relations to reduce the six-point disk integral $Z(12|34|56)$, where 
 \bea
Z(12|34|56)\equiv 
 \int \limits_{0\leq  z_{2} \leq z_{3}\leq z_{5}\leq 1 }
 \left(\prod_{i=2}^5 d^2 z_i\right)\, \frac{{\rm KN} }{z^2_{12}z^{2}_{34}z^{2}_{56}} \, , 
\eea
and ${\rm KN}=\prod^n_{i<l} |z_{il}|^{s_{il}}$, to linear combinations of ``single trace" integrals $Z(123456)$.\footnote{From now on we will suppress the notation for integration regions, knowing that we always have $0\leq  z_{2} \leq z_{3}\leq z_{5}\leq 1$.} For example, consider $Z(23|1456)$ where 
\eq
Z(23|1465)\equiv \int 
 \left(\prod_{i=2}^5 d^2 z_i\right) \frac{{\rm KN}}{z^2_{23}z_{14}z_{46}z_{65}z_{51}} =-\int 
 \left(\prod_{i=2}^5 d^2 z_i\right)\, \frac{{\rm KN} }{z_{14}z_{46}z_{65}z_{51}}\p_{z_2}\frac{1}{z_{23}}\,.
\eqe
Integrating by parts while keeping in mind that $z_6$ has been set to $\infty$, one finds:
\eq
-\int \frac{{\rm KN} }{z_{14}z_{51}}\p_{z_2}\frac{1}{z_{23}}=\int \frac{{\rm KN} }{z_{23}z_{14}z_{51}}\left(\frac{s_{21}}{z_{21}}+\frac{s_{23}}{z_{23}}+\frac{s_{24}}{z_{24}}+\frac{s_{25}}{z_{25}}\right)\,.
\eqe
In the above we have surpressed terms involving $z_6$, which can be restored simply by ensuring one has the correct SL(2) weights at each point, i.e.
\eq
-\int \frac{{\rm KN} }{z_{14}z_{46}z_{65}z_{51}}\p_{z_2}\frac{1}{z_{23}}=\int \frac{{\rm KN} }{z_{14}z_{51}z_{23}}\left(\frac{s_{23}}{z_{23}z_{46}z_{65}}+\frac{s_{24}}{z_{24}z_{36}z_{65}}+\frac{s_{25}}{z_{25}z_{46}z_{63}}\right)\,.
\eqe
This leads to the identity 
\eq\label{Z24Reduce}
Z(23|1465)=\left(s_{12}(Z(145632)+Z(154632))-s_{24}Z(142365)-s_{23}Z(146325)\right)/(1-s_{23})\,.
\eqe
Similarly one has:
\eqa\label{Z33Reduce}
Z(123|456)=\left(s_{34}Z(621345)-s_{35}Z(621354)+s_{15}Z(154623)-s_{14}Z(145623)\right)/(1-s_{123})\nonumber\\\,.
\eqae
Finally, repeated use of IBP relations lead to 
\eqa
Z(23|14|56)=s_{25}Z(14|2365) +&&\left[\frac{s_{12}}{s_{14}-1}
   \left(s_{24}(A-Z(214|536))-s_{34}Z(56|1432)+s_{45}Z(321456) \right)\right. \nonumber\\
&&\left. \quad \quad+1\leftrightarrow4 \right]\frac{1}{1-s_{23}}
\eqae
where 
\eq
A=\frac{1}{s_{56}-1}\left(s_{16}Z(614235)-s_{64}Z(356412)+s_{36}Z(563|142)\right)\,.
\eqe
Applying the result in eq.~(\ref{Z24Reduce}) and  eq.~(\ref{Z33Reduce}), one recovers a result that is expressed in terms of single trace integrands.

\bibliographystyle{JHEP}
\bibliography{softref}{}

\providecommand{\href}[2]{#2}\begingroup\raggedright\begin{thebibliography}{10}

\bibitem{Weinberg:1965nx}
S.~Weinberg, \emph{{Infrared photons and gravitons}},
  \href{http://dx.doi.org/10.1103/PhysRev.140.B516}{\emph{Phys. Rev.} {\bf 140}
  (1965) B516--B524}.

\bibitem{Adler:1964um}
S.~L. Adler, \emph{{Consistency conditions on the strong interactions implied
  by a partially conserved axial vector current}},
  \href{http://dx.doi.org/10.1103/PhysRev.137.B1022}{\emph{Phys. Rev.} {\bf
  137} (1965) B1022--B1033}.

\bibitem{Low:1958sn}
F.~Low, \emph{{Bremsstrahlung of very low-energy quanta in elementary particle
  collisions}},
  \href{http://dx.doi.org/10.1103/PhysRev.110.974}{\emph{Phys.Rev.} {\bf 110}
  (1958) 974--977}.

\bibitem{Burnett:1967km}
T.~Burnett and N.~M. Kroll, \emph{{Extension of the low soft photon theorem}},
  \href{http://dx.doi.org/10.1103/PhysRevLett.20.86}{\emph{Phys.Rev.Lett.} {\bf
  20} (1968) 86}.

\bibitem{GellMann:1954kc}
M.~Gell-Mann and M.~L. Goldberger, \emph{{Scattering of low-energy photons by
  particles of spin 1/2}},
  \href{http://dx.doi.org/10.1103/PhysRev.96.1433}{\emph{Phys. Rev.} {\bf 96}
  (1954) 1433--1438}.

\bibitem{Gross:1968in}
D.~J. Gross and R.~Jackiw, \emph{{Low-Energy Theorem for Graviton Scattering}},
  \href{http://dx.doi.org/10.1103/PhysRev.166.1287}{\emph{Phys.Rev.} {\bf 166}
  (1968) 1287--1292}.

\bibitem{Bern:2014vva}
Z.~Bern, S.~Davies, P.~Di~Vecchia and J.~Nohle, \emph{{Low-Energy Behavior of
  Gluons and Gravitons from Gauge Invariance}},
  \href{https://arxiv.org/abs/1406.6987}{{\tt 1406.6987}}.

\bibitem{Low:2015ogb}
I.~Low, \emph{{Double Soft Theorems and Shift Symmetry in Nonlinear Sigma
  Models}}, \href{http://dx.doi.org/10.1103/PhysRevD.93.045032}{\emph{Phys.
  Rev.} {\bf D93} (2016) 045032}, [\href{https://arxiv.org/abs/1512.01232}{{\tt
  1512.01232}}].

\bibitem{Cheung:2015ota}
C.~Cheung, K.~Kampf, J.~Novotny, C.-H. Shen and J.~Trnka, \emph{{On-Shell
  Recursion Relations for Effective Field Theories}},
  \href{http://dx.doi.org/10.1103/PhysRevLett.116.041601}{\emph{Phys. Rev.
  Lett.} {\bf 116} (2016) 041601},
  [\href{https://arxiv.org/abs/1509.03309}{{\tt 1509.03309}}].

\bibitem{Luo:2015tat}
H.~Luo and C.~Wen, \emph{{Recursion relations from soft theorems}},
  \href{http://dx.doi.org/10.1007/JHEP03(2016)088}{\emph{JHEP} {\bf 03} (2016)
  088}, [\href{https://arxiv.org/abs/1512.06801}{{\tt 1512.06801}}].

\bibitem{Arkani-Hamed:2016rak}
N.~Arkani-Hamed, L.~Rodina and J.~Trnka, \emph{{Locality and Unitarity from
  Singularities and Gauge Invariance}},
  \href{https://arxiv.org/abs/1612.02797}{{\tt 1612.02797}}.

\bibitem{Rodina:2016mbk}
L.~Rodina, \emph{{Uniqueness from locality and BCFW shifts}},
  \href{https://arxiv.org/abs/1612.03885}{{\tt 1612.03885}}.

\bibitem{Rodina:2016jyz}
L.~Rodina, \emph{{Uniqueness from gauge invariance and the Adler zero}},
  \href{https://arxiv.org/abs/1612.06342}{{\tt 1612.06342}}.

\bibitem{Bianchi:2016viy}
M.~Bianchi, A.~L. Guerrieri, Y.-t. Huang, C.-J. Lee and C.~Wen,
  \emph{{Exploring soft constraints on effective actions}},
  \href{http://dx.doi.org/10.1007/JHEP10(2016)036}{\emph{JHEP} {\bf 10} (2016)
  036}, [\href{https://arxiv.org/abs/1605.08697}{{\tt 1605.08697}}].

\bibitem{Cheung:2016drk}
C.~Cheung, K.~Kampf, J.~Novotny, C.-H. Shen and J.~Trnka, \emph{{A Periodic
  Table of Effective Field Theories}},
  \href{http://dx.doi.org/10.1007/JHEP02(2017)020}{\emph{JHEP} {\bf 02} (2017)
  020}, [\href{https://arxiv.org/abs/1611.03137}{{\tt 1611.03137}}].

\bibitem{Chen:2015hpa}
W.-M. Chen, Y.-t. Huang and C.~Wen, \emph{{Exact coefficients for higher
  dimensional operators with sixteen supersymmetries}},
  \href{http://dx.doi.org/10.1007/JHEP09(2015)098}{\emph{JHEP} {\bf 09} (2015)
  098}, [\href{https://arxiv.org/abs/1505.07093}{{\tt 1505.07093}}].

\bibitem{Yennie:1961ad} 
  D.~R.~Yennie, S.~C.~Frautschi and H.~Suura,
  Annals Phys.\  {\bf 13}, 379 (1961).
  doi:10.1016/0003-4916(61)90151-8

\bibitem{Laenen:2010uz} 
  E.~Laenen, L.~Magnea, G.~Stavenga and C.~D.~White,
  JHEP {\bf 1101}, 141 (2011)
  doi:10.1007/JHEP01(2011)141
  [arXiv:1010.1860 [hep-ph]].


\bibitem{Low:2001bw}
I.~Low and A.~V. Manohar, \emph{{Spontaneously broken space-time symmetries and
  Goldstone's theorem}},
  \href{http://dx.doi.org/10.1103/PhysRevLett.88.101602}{\emph{Phys. Rev.
  Lett.} {\bf 88} (2002) 101602},
  [\href{https://arxiv.org/abs/hep-th/0110285}{{\tt hep-th/0110285}}].

\bibitem{DiVecchia:2015jaq}
P.~Di~Vecchia, R.~Marotta, M.~Mojaza and J.~Nohle, \emph{{New soft theorems for
  the gravity dilaton and the Nambu-Goldstone dilaton at subsubleading order}},
  \href{http://dx.doi.org/10.1103/PhysRevD.93.085015}{\emph{Phys. Rev.} {\bf
  D93} (2016) 085015}, [\href{https://arxiv.org/abs/1512.03316}{{\tt
  1512.03316}}].

\bibitem{He:2014bga}
S.~He, Y.-t. Huang and C.~Wen, \emph{{Loop Corrections to Soft Theorems in
  Gauge Theories and Gravity}},
  \href{http://dx.doi.org/10.1007/JHEP12(2014)115}{\emph{JHEP} {\bf 12} (2014)
  115}, [\href{https://arxiv.org/abs/1405.1410}{{\tt 1405.1410}}].

\bibitem{Bianchi:2014gla}
M.~Bianchi, S.~He, Y.-t. Huang and C.~Wen, \emph{{More on Soft Theorems: Trees,
  Loops and Strings}},
  \href{http://dx.doi.org/10.1103/PhysRevD.92.065022}{\emph{Phys. Rev.} {\bf
  D92} (2015) 065022}, [\href{https://arxiv.org/abs/1406.5155}{{\tt
  1406.5155}}].

\bibitem{Elvang:2016qvq}
H.~Elvang, C.~R.~T. Jones and S.~G. Naculich, \emph{{Soft Photon and Graviton
  Theorems in Effective Field Theory}},
  \href{https://arxiv.org/abs/1611.07534}{{\tt 1611.07534}}.

\bibitem{Cachazo:2015ksa}
F.~Cachazo, S.~He and E.~Y. Yuan, \emph{{New Double Soft Emission Theorems}},
  \href{https://arxiv.org/abs/1503.04816}{{\tt 1503.04816}}.

\bibitem{ArkaniHamed:2008gz}
N.~Arkani-Hamed, F.~Cachazo and J.~Kaplan, \emph{{What is the Simplest Quantum
  Field Theory?}}, \href{http://dx.doi.org/10.1007/JHEP09(2010)016}{\emph{JHEP}
  {\bf 1009} (2010) 016}, [\href{https://arxiv.org/abs/0808.1446}{{\tt
  0808.1446}}].

\bibitem{Elvang:2010kc}
H.~Elvang and M.~Kiermaier, \emph{{Stringy KLT relations, global symmetries,
  and $E_{7(7)}$ violation}},
  \href{http://dx.doi.org/10.1007/JHEP10(2010)108}{\emph{JHEP} {\bf 10} (2010)
  108}, [\href{https://arxiv.org/abs/1007.4813}{{\tt 1007.4813}}].

\bibitem{Beisert:2010jx}
N.~Beisert, H.~Elvang, D.~Z. Freedman, M.~Kiermaier, A.~Morales and
  S.~Stieberger, \emph{{E7(7) constraints on counterterms in N=8
  supergravity}},
  \href{http://dx.doi.org/10.1016/j.physletb.2010.09.069}{\emph{Phys. Lett.}
  {\bf B694} (2011) 265--271}, [\href{https://arxiv.org/abs/1009.1643}{{\tt
  1009.1643}}].

\bibitem{Volkov:1972jx}
D.~V. Volkov and V.~P. Akulov, \emph{{Possible universal neutrino
  interaction}}, {\emph{JETP Lett.} {\bf 16} (1972) 438--440}.

\bibitem{Volkov:1973ix}
D.~V. Volkov and V.~P. Akulov, \emph{{Is the Neutrino a Goldstone Particle?}},
  \href{http://dx.doi.org/10.1016/0370-2693(73)90490-5}{\emph{Phys. Lett.} {\bf
  46B} (1973) 109--110}.

\bibitem{DiVecchia:2017uqn}
P.~Di~Vecchia, R.~Marotta and M.~Mojaza, \emph{{Double-soft behavior of the
  dilaton of spontaneously broken conformal invariance}},
  \href{https://arxiv.org/abs/1705.06175}{{\tt 1705.06175}}.

\bibitem{Boels:2015pta}
R.~H. Boels and W.~Wormsbecher, \emph{{Spontaneously broken conformal
  invariance in observables}},  \href{https://arxiv.org/abs/1507.08162}{{\tt
  1507.08162}}.

\bibitem{Huang:2015sla}
Y.-t. Huang and C.~Wen, \emph{{Soft theorems from anomalous symmetries}},
  \href{http://dx.doi.org/10.1007/JHEP12(2015)143}{\emph{JHEP} {\bf 12} (2015)
  143}, [\href{https://arxiv.org/abs/1509.07840}{{\tt 1509.07840}}].

\bibitem{Bianchi:2015cta}
M.~Bianchi, J.~F. Morales and C.~Wen, \emph{{Instanton corrections to the
  effective action of $ \mathcal{N}=4 $ SYM}},
  \href{http://dx.doi.org/10.1007/JHEP11(2015)006}{\emph{JHEP} {\bf 11} (2015)
  006}, [\href{https://arxiv.org/abs/1508.00554}{{\tt 1508.00554}}].

\bibitem{Kuzenko:2005wh}
S.~M. Kuzenko and S.~A. McCarthy, \emph{{On the component structure of N=1
  supersymmetric nonlinear electrodynamics}},
  \href{http://dx.doi.org/10.1088/1126-6708/2005/05/012}{\emph{JHEP} {\bf 05}
  (2005) 012}, [\href{https://arxiv.org/abs/hep-th/0501172}{{\tt
  hep-th/0501172}}].

\bibitem{deWit:1975xci}
B.~de~Wit and D.~Z. Freedman, \emph{{Phenomenology of Goldstone Neutrinos}},
  \href{http://dx.doi.org/10.1103/PhysRevLett.35.827}{\emph{Phys. Rev. Lett.}
  {\bf 35} (1975) 827}.

\bibitem{Chen:2014xoa}
W.-M. Chen, Y.-t. Huang and C.~Wen, \emph{{New Fermionic Soft Theorems for
  Supergravity Amplitudes}},
  \href{http://dx.doi.org/10.1103/PhysRevLett.115.021603}{\emph{Phys. Rev.
  Lett.} {\bf 115} (2015) 021603}, [\href{https://arxiv.org/abs/1412.1809}{{\tt
  1412.1809}}].

\bibitem{Kallosh:2016qvo}
R.~Kallosh, \emph{{Nonlinear (Super)Symmetries and Amplitudes}},
  \href{http://dx.doi.org/10.1007/JHEP03(2017)038}{\emph{JHEP} {\bf 03} (2017)
  038}, [\href{https://arxiv.org/abs/1609.09123}{{\tt 1609.09123}}].

\bibitem{Kallosh:2016lwj}
R.~Kallosh, A.~Karlsson and D.~Murli, \emph{{Origin of Soft Limits from
  Nonlinear Supersymmetry in Volkov-Akulov Theory}},
  \href{http://dx.doi.org/10.1007/JHEP03(2017)081}{\emph{JHEP} {\bf 03} (2017)
  081}, [\href{https://arxiv.org/abs/1609.09127}{{\tt 1609.09127}}].
  
\bibitem{Komargodski:2009rz}
Z.~Komargodski and N.~Seiberg, \emph{{From Linear SUSY to Constrained
  Superfields}},
  \href{http://dx.doi.org/10.1088/1126-6708/2009/09/066}{\emph{JHEP} {\bf 09}
  (2009) 066}, [\href{https://arxiv.org/abs/0907.2441}{{\tt 0907.2441}}].

\bibitem{Kuzenko:2010ef}
S.~M. Kuzenko and S.~J. Tyler, \emph{{Relating the Komargodski-Seiberg and
  Akulov-Volkov actions: Exact nonlinear field redefinition}},
  \href{http://dx.doi.org/10.1016/j.physletb.2011.03.020}{\emph{Phys. Lett.}
  {\bf B698} (2011) 319--322}, [\href{https://arxiv.org/abs/1009.3298}{{\tt
  1009.3298}}].

\bibitem{Huang:2016tag}
Y.-t. Huang, O.~Schlotterer and C.~Wen, \emph{{Universality in string
  interactions}}, \href{http://dx.doi.org/10.1007/JHEP09(2016)155}{\emph{JHEP}
  {\bf 09} (2016) 155}, [\href{https://arxiv.org/abs/1602.01674}{{\tt
  1602.01674}}].

\bibitem{Mafra:2011nv}
C.~R. Mafra, O.~Schlotterer and S.~Stieberger, \emph{{Complete N-Point
  Superstring Disk Amplitude I. Pure Spinor Computation}},
  \href{http://dx.doi.org/10.1016/j.nuclphysb.2013.04.023}{\emph{Nucl. Phys.}
  {\bf B873} (2013) 419--460}, [\href{https://arxiv.org/abs/1106.2645}{{\tt
  1106.2645}}].

\bibitem{Schlotterer:2012ny}
O.~Schlotterer and S.~Stieberger, \emph{{Motivic Multiple Zeta Values and
  Superstring Amplitudes}},
  \href{http://dx.doi.org/10.1088/1751-8113/46/47/475401}{\emph{J. Phys.} {\bf
  A46} (2013) 475401}, [\href{https://arxiv.org/abs/1205.1516}{{\tt
  1205.1516}}].

\bibitem{OliverWeb}
J.~Broedel, O.~Schlotterer and S.~Stieberger, \emph{{\tt
  http://mzv.mpp.mpg.de}}, .

\bibitem{Broedel:2013tta}
J.~Broedel, O.~Schlotterer and S.~Stieberger, \emph{{Polylogarithms, Multiple
  Zeta Values and Superstring Amplitudes}},
  \href{http://dx.doi.org/10.1002/prop.201300019}{\emph{Fortsch. Phys.} {\bf
  61} (2013) 812--870}, [\href{https://arxiv.org/abs/1304.7267}{{\tt
  1304.7267}}].

\bibitem{Cheung:2017ems}
C.~Cheung, C.-H. Shen and C.~Wen, \emph{{Unifying Relations for Scattering
  Amplitudes}},  \href{https://arxiv.org/abs/1705.03025}{{\tt 1705.03025}}.

\end{thebibliography}\endgroup

\end{document}